\documentstyle[psfig]{article}

\newcommand{\param}{\theta}
\newcommand{\diffAngle}{\xi}
\newcommand{\varDet}{x}
\newcommand{\varPupil}{{x_{{\rm p}}}}
\newcommand{\xio}{\xi_{0}}
\newcommand{\p}{p}              
\newcommand{\cond}{\,\vert\,}   
\newcommand{\D}[2]{\frac{\partial{#1}}{\partial{#2}}}  
\newcommand{\Cramer}{Cram\'er-Rao}
\newcommand{\EFI}{Fisher information}
\newcommand{\FID}{Fisher information density}
\newcommand{\FIDs}{Fisher information densities}
\newcommand{\PDF}{likelihood function}          
\newcommand{\PDFs}{likelihood functions}          
\newcommand{\varFID}{i_{{\rm F}}} 
\newcommand{\halfSegWidth}{w} 
\newcommand{\centerSegment}{c} 
\newcommand{\boxheight}{relative segment phase}
\newcommand{\boxheights}{differential segment phases}
\newcommand{\Boxheights}{Differential segment phases}
\newcommand{\stepheight}{phase step}
\newcommand{\stepheights}{phase steps}
\newcommand{\complAmpl}{U}

\begin{document}
\title{Optical measurements of phase steps in segmented mirrors -
 fundamental precision limits}

\author{L.~Noethe$^{a}$ and H.-M.~Adorf$^{b}$
\vspace*{3mm}\\
$^{a}$ European Southern Observatory, Garching bei M\"{u}nchen, Germany\\
$^{b}$ Max-Planck-Institut f\"{u}r Astrophysik, Garching bei M\"{u}nchen, Germany}
\date{}
\maketitle

\noindent {\bf Abstract}
\noindent Phase steps are an important type of wavefront aberrations generated by
large telescopes with segmented mirrors. In a closed-loop correction cycle these
phase steps have to be measured with the highest possible precision using natural
reference stars, that is with a small number of photons. In this paper the classical
Fisher information of statistics is used for calculating the \Cramer\ bound, which
determines the limit to the precision with which the height of the steps can be
estimated in an unbiased fashion with a given number of photons and a given
measuring device. Four types of measurement devices are discussed: a Shack-Hartmann
sensor with one small cylindrical lenslet covering a sub-aperture centred over a
border, a modified Mach-Zehnder interferometer, a Foucault test, and a curvature sensor.
The \Cramer\ bound is calculated for all sensors under ideal conditions, that is
narrowband measurements without additional noise or disturbances apart from the
photon shot noise.
This limit is compared with the ultimate quantum statistical limit for the estimate of
such a step which is independent of the measuring device.
For the Shack-Hartmann sensor, the effects on the \Cramer\ bound of broadband measurements,
finite sampling, and disturbances such as atmospheric seeing and detector readout noise
are also investigated.
The methods presented here can be used to compare the precision limits of various
devices for measuring \boxheights\ and for optimising the parameters of the
devices. Under ideal conditions the Shack-Hartmann and the Foucault devices nearly
attain the ultimate quantum statistical limits, whereas the Mach-Zehnder and the
curvature devices each require approximately twenty times as many photons
in order to reach the same precision.

\section{Introduction}
In future extremely large telescopes at least one of the mirrors will probably
be segmented. During the operation of such a telescope the wavefront errors
introduced by segmentation have to be reduced to the order of a few tens of nanometers.
If the segments are regarded as rigid bodies, the wavefront errors can only
be generated by tip-tilt errors of individual segments or differential
axial displacements.
This paper assumes that the segments are already perfectly aligned in tip and tilt.
A correction of remaining piston errors -- from now on called \boxheights\ --
requires measurements of the heights of the wavefront steps at the borders
of adjacent segments with a precision of a few nanometers,
and the suppression of the \boxheight s by appropriate
rigid body movements of the segments.

Several methods are used or have been proposed for the measurement
of \boxheights\ in segmented mirrors. One method is based on the
Shack-Hartmann technique and applied in a slightly
modified fashion at the Keck
telescope~\cite{Chanan1998}\cite{Chanan2000}\cite{Chanan2004}.
In an image of the segmented mirror a circular lens
is placed over a sub-aperture centred on an intersegment border.
The \stepheight\ across the border is estimated from the form of the diffraction
pattern, for narrowband and broadband measurements and in
the presence of atmospheric disturbances. Other
methods proposed in the literature use spatial filtering in a modified
Mach-Zehnder interferometer~\cite{Yaitskova2005},
measurements in defocussed images and pupils~\cite{Roddier1987}
\cite{Chanan1999} \cite{Gonzales2001} \cite{Schumacher2005},
phase filtering techniques~\cite{Dohlen2004}, or a knife
edge technique in the pyramid wavefront
sensor~\cite{Esposito2001}\cite{Esposito2003}\cite{Pinna2004}\cite{Esposito2005}.

All the sensors discussed in this paper generate diffraction patterns in the
plane of a detector. These can be regarded as likelihood functions 
$\p(\varDet \cond \param)$ describing the ``likelihood'' to register a photon
on the detector as a function of the spatial variable $\varDet$ on the detector,
given a certain value of a parameter, in this paper the \boxheight\ $\param$.
There are various possibilities to define,
based on the derivatives of the $\p(\varDet \cond \param)$
with respect to $\param$, an information
content in the \PDF\. One of them is the classical \EFI\ 
\cite{KendallStuart1979}:
\begin{equation}
   I_{{\rm F}} = \int_{-\infty}^{+\infty} {\rm d}\varDet \, \, p(\varDet \cond \param) 
            \left( \frac{\partial \ln p(\varDet \cond \param)}{\partial \param} \right)^{2}
            \, .
   \label{eq:fisherInformationGeneral}
\end{equation}
Estimating the \boxheights\ is particularly difficult at low
light levels, e.g.\ when only the light of faint stars is available
for the measurements. The question arises what precision
can ultimately be obtained with a given number of detected photons.
In statistics, precision is characterized by the conditional
variance of a quantity. For a parameter of interest,
in this paper the \boxheight~$\param$, the
minimum variance of an unbiased estimate of the parameter,
the so-called \Cramer\ minimum
variance bound~$\sigma^2_{\rm CRB}$~\cite{KendallStuart1979},
is given by the inverse of
the classical Fisher information~$I_{{\rm F}}(\theta)$:
\begin{equation}
   \sigma^{2}_{\rm CRB}(\theta) = 1 / I_{{\rm F}}(\theta)
   \quad.
   \label{eq:classicalFisherInfo}
\end{equation}
In the following the shorter expression \Cramer\ bound will be used for the
square root $\sigma_{{\rm CRB}}$ of the \Cramer\ minimum
variance bound.

Two of the assumptions in the derivation of equation~(\ref{eq:classicalFisherInfo})
are, unless it is obvious, checked for all cases investigated in this paper.
First, the integral of the \PDF\ over the
interval for which it is defined must not depend on $\param$, at least for a given value of
$\param$ for which the \EFI\ is calculated. Second, the \PDF\
must be regular with respect to $\param$.
Under the first assumption the \PDF\ can be normalised:
\begin{equation}
   \int_{-\infty}^{+\infty}
            {\rm d}\varDet \, p(\varDet \cond \param) = 1.
   \label{eq:integralPDF}
\end{equation}
This is equivalent to the assumption that one photon passes through the aperture.

If $q(\varDet \cond \param)$ denotes the real-valued
likelihood amplitude, defined by
\begin{equation}
   q(\varDet \cond \param) = \sqrt{p(\varDet \cond \param)},
   \label{eq:defProbAmplitude}
\end{equation}
the integrand in equation (\ref{eq:fisherInformationGeneral}), which we will call the
{\it Fisher information density} $\varFID(\varDet\cond\param)$, can be written as
\begin{equation}
   \varFID(\varDet,\param)
   = 4\, \left(\D{q(\varDet \cond \param)}{\param} \right)^{2}.
   \label{eq:defFisherDensity}
\end{equation}

In principle there is an ambiguity related to the definition of the \FID.
This is due the identity
\begin{equation}
   \int_{-\infty}^{+\infty} \, {\rm d}\varDet \,
             \left(\D{q(\varDet \cond \param)}{\param} \right)^{2} =
   - \int_{-\infty}^{+\infty} \, {\rm d}\varDet \,
               q(\varDet \cond \param) \, 
       \frac{\partial^{2}q(\varDet \cond \param)}{\partial \param^{2}}
   \label{eq:equalityFIDs}
\end{equation}
which can be obtained by differentiating equation~(\ref{eq:integralPDF})
twice with respect to $\param$.
The Fisher information defined as
\begin{equation}
   I_{{\rm F}} = 4\, \int_{-\infty}^{+\infty} \, {\rm d}\varDet \,
                \left[
                 A\, \left(\D{q(\varDet \cond \param)}{\param} \right)^{2} -
                 B\, q(\varDet \cond \param) \, 
                      \frac{\partial^{2}q(\varDet \cond \param)}{\partial \param^{2}}
                \right]
   \label{eq:possibleFID}
\end{equation}
would be independent of the choice of $A$ and $B$ provided that $A + B = 1$.
However, the term following the coefficient $B$ can be negative, possibly giving rise
to a negative \FID. Since this seems unreasonable, $B$ is set to zero
and the \FID\ is defined as in equation~(\ref{eq:defFisherDensity}).
The \EFI\ will therefore be computed as
\begin{equation}
   I_{{\rm F}} = 4\, \int_{-\infty}^{+\infty} \, {\rm d}\varDet \,
                  \left(\D{q(\varDet \cond \param)}{\param} \right)^{2}.
   \label{eq:EFI}
\end{equation}
According to equation~(\ref{eq:classicalFisherInfo}) the computation of the Fisher information
offers a possibility to calculate analytically the
potential precision limits of various measurement methods.
These limits can then be compared with the ultimate limit which is independent of any
measuring device.
The latter is calculated in section~\ref{sec:ultimateLimit} from first principles
by introducing the wave function describing the wavefront immediately after the reflection
by the mirror, generating a \boxheight, into the quantum statistical counterpart of the
expressions~(\ref{eq:fisherInformationGeneral})
or (\ref{eq:EFI}) for the Fisher information. 

Section~\ref{sec:CRBsSensorsIdeal} discusses the \Cramer\ bounds for the Shack-Hartmann,
the Mach-Zehnder, the Foucault and the curvature sensor under ideal conditions, that is
for monochromatic light and without any effects introducing additional noise or light losses
or aberrations. 
The effects of broadband measurements, finite sampling, and external disturbances
such as seeing and detector readout noise can also be included in the calculations.
The methods are outlined in detail only for the Shack-Hartmann sensor
in section~\ref{sec:noiseEffects},
but could readily be applied to the other sensors as well.

In general, the limits given by the \Cramer\ bound can only be reached
if efficient estimators are available for the analysis of the data. This issue
will, however, not be discussed in this paper. Furthermore, all calculations
will be carried out for the one-dimensional case.

\section{Definition of the wavefront error}
\begin{figure}[h]
  \centerline{\hbox{
   \psfig{figure=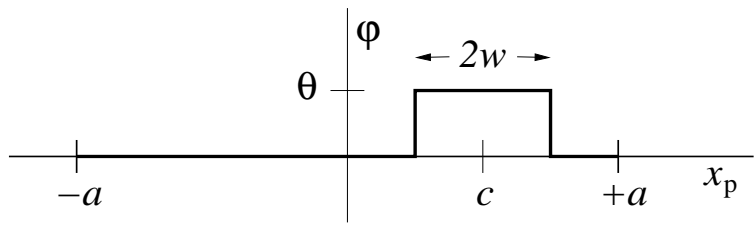,width=100mm}}}
   \caption{\label{fig:step} {\small Wavefront phase $\varphi$ with
     a relative segment phase~$\param$ generated by a segment of
      width $2\halfSegWidth$ and a centre at $\centerSegment$
      within a subaperture of width~$2a$.}}
\end{figure}

Let, as shown in figure~\ref{fig:step}, the pupil be defined by the interval $[-a,+a]$
with the spatial variable in the pupil denoted by $\varPupil$.
An out-of-phase segment with its centre at $\centerSegment$
generates a constant wavefront error or phase difference
$\varphi(\varPupil)=\param$ over its full width
$2\halfSegWidth$ compared to a phase, also constant, over the rest of the
pupil.
There are two wavefront discontinuities, that is phase steps 
of height~$\param$, at the segment borders. Apart from an irrelevant average phase,
the complex amplitude describing the wavefront is given by
\begin{equation}
  \complAmpl_{{\rm s}}(\varPupil \cond \param) = \frac{1}{\sqrt{2a}}
        \, e^{i k \varphi(\varPupil)} \; W(\varPupil,-a,+a)
     \quad, \label{eq:defPsiS}
\end{equation}
with
\begin{equation}
  \varphi(\varPupil) =
    \left\{
        \begin{array}{lll}
	0 & {\rm for} & -a \le \varPupil \le \centerSegment - \halfSegWidth\\
        \param & {\rm for} & \centerSegment - \halfSegWidth < \varPupil < \centerSegment + \halfSegWidth\\
        0 & {\rm for} & \centerSegment + \halfSegWidth \le \varPupil \le a
        \end{array}
    \right\}
    \label{eq:defVarphi}
\end{equation}
and
\begin{equation}
W(\varPupil,-a,+a) = 
 \left\{
        \begin{array}{lll}
	0 & {\rm for} & \varPupil < -a\\
        1 & {\rm for} & -a \le \varPupil \le +a\\
        0 & {\rm for} & +a < \varPupil
        \end{array}
    \right\}
    \quad, \label{eq:defBoxcar}
\end{equation}
where $k = 2\pi/ \lambda$ is the wavenumber and $\lambda$ is the
wavelength of the light.
The restriction of the definition to the interval $[-a, +a]$
signifies that only the photons passing through the aperture of
width~$2a$ will be used for estimating the \boxheight\ $ßparam$,
and the factor in front of the exponential in equation~(\ref{eq:defPsiS})
normalizes the wave function such that the number of photons within
this sub-aperture is equal to~1, that is
\begin{equation}
  \int_{-a}^{+a} d\varPupil \, \complAmpl_{{\rm s}}^{\ast}(\varPupil \cond \param)
          \, \complAmpl_{{\rm s}}(\varPupil \cond \param) = 1
    \quad,
    \label{eq:normalisationU}
\end{equation}
where $\complAmpl_{{\rm s}}^{\ast}$ is the complex conjugate of $\complAmpl_{{\rm s}}$.
Apart from the general or two-border configuration shown in figure~\ref{fig:step},
two special configurations will be discussed in more detail. One is given if the segment
is located at the edge of the aperture $[-a,+a]$, that is if $c=a-\halfSegWidth$. This
configuration will subsequently be called the {\it single-border} configuration. The second special
configuration, which is a special case of the single-border configuration, is given
if the border is at the centre of the aperture $[-a,+a]$, that is if
$c=\halfSegWidth=a/2$. This configuration, which is of particular interest for the
Shack-Hartmann type sensor, will be called the {\it centred single-border} configuration.
For the general configuration the parameter~$\param$ will be referred to as the
\boxheight, and for the single-border configurations as the \stepheight. 

\section{Quantum statistical precision limit for the estimate of a \boxheight}
\label{sec:ultimateLimit}
The ultimate precision limit for the measurement of \boxheights\ can, if at all, be
obtained with a measuring device which is capable of transferring
the maximum relevant information contained in the source to the data. In addition,
one needs an appropriate, that is efficient data analysis procedure.

As stated in the introduction, the limit on the data side can be
obtained from the Fisher information which is derived from a
classical, real-valued \PDF.
However, on the source side, immediately after the
reflection of an incoming monochromatic plane wavefront by the mirror configuration
shown in figure~\ref{fig:step}, the full information is contained
in a complex quantum statistical wave function $\psi(x \cond \param)$,
which is equivalent to the complex amplitude function~(\ref{eq:defPsiS})
familiar from optical theory.
Therefore, the calculation of the information
in the source requires the quantum statistical equivalent
of the expression~(\ref{eq:EFI})
for the Fisher information. For a pure state $\psi$ and
its complex conjugate $\psi^{\ast}$ this is given \cite{BraunsteinCavesMilburn1996} by
\begin{equation}
    I_{{\rm QM}}(\param) = 4\, \left( \int_{-a}^{+a} {\rm d}\varPupil
      \D{}{\param} \psi^{\ast}(\varPupil\cond\param)
      \D{}{\param} \psi(\varPupil\cond\param)\; - \;
       \left| \int_{-a}^{+a} {\rm d}\varPupil \;
            \psi^{\ast}(\varPupil\cond\param)
            \D{}{\param} \psi(\varPupil\cond\param)
    \right|^{2} \right)
    \quad .
    \label{eq:informationSource}
\end{equation}
The first term on the right hand side is similar to the definition
of the classical \FID\ in equation~(\ref{eq:EFI}).
The second term, in effect, suppresses a dependence
of $I_{{\rm QM}}$ on an undetectable average phase across the full aperture $[-a,+a]$.

By replacing the complex wave function $\psi$
in equation~(\ref{eq:informationSource}) by the
complex amplitude $\complAmpl_{{\rm s}}$ of equation~(\ref{eq:defPsiS}) one obtains
the quantum statistical equivalent of the Fisher information about the
\boxheight\ contained in the complex amplitude immediately after the reflection
by the mirrors:
\begin{equation}
  I_{{\rm QM}}(\param) = 4\, k^{2} \,
         \frac{\halfSegWidth}{a} \,
   \left( 1 - \frac{\halfSegWidth}{a} \right)
  \quad.
  \label{eq:informationSourcePiston}
\end{equation}
The quadratic dependence of the quantum statistical information about the
segment width, which is independent of the segment position,
is shown in figure~\ref{fig:efiCompareSensors} by the solid line. The maximum
\begin{equation}
  I_{{\rm QM,max}}(\param) = k^{2}
           \label{eq:efiQMmax}
\end{equation}
is obtained if the segment covers half of the aperture, that is for $w=a/2$.

For a single photon, the lower bound for the root-mean-square (RMS)
of the uncertainty in the estimate of the parameter $\param$,
that is the quantum statistical equivalent of the square of the classical
\Cramer\ minimum variance bound, is given by
\begin{equation}
  \sigma_{{\rm QM,CRB},1}(\param) = \frac{1}{2k} \frac{a}{\sqrt{w(a-w)}}
  \quad.
  \label{eq:sigmaLimit}
\end{equation}
The minimum value is obtained, if the segment covers half of the pupil:
\begin{equation}
  \sigma_{{\rm QM,CRB},1,{\rm min}}(\param) = \frac{1}{k} = \frac{\lambda}{2\pi}
  \quad.
  \label{eq:sigmaLimitMinimum}
\end{equation}
It depends only on the wavelength of the light, and is
of the order of $100$~nm for visible wavelengths.
For~$N_{\rm phot}$ registered photons the \EFI\ increases by a
factor of~$N_{\rm phot}$, the quantum statistical \Cramer\ bound
decreases by $1/\sqrt{N_{\rm phot}}$, and the lower bound for the RMS
of the uncertainty in the estimate
of the parameter $\param$ is correspondingly given by
\begin{equation}
  \sigma_{\rm QM,CRB,{\rm min}}(\param) =
           \frac{1}{\sqrt{N_{\rm phot}}} \, \frac{\lambda}{2\pi}
  \quad.
     \label{eq:sigmaLimitNPhot}
\end{equation}
This fundamental limit does
not refer to a particular measurement device or a particular data analysis
procedure. It is based on the maximum information that can potentially be
extracted from the wavefront, without necessarily implying that such an efficient
measurement process exists.

\section{\Cramer\ bounds for the measurement of piston steps with different sensors
 under ideal conditions}
\label{sec:CRBsSensorsIdeal}
The \Cramer\ bounds for measuring the \boxheight\ or \stepheight~$\param$  are calculated
for the four sensors
mentioned above, under ideal conditions, that is under the following assumptions:
light from a quasi-monochromatic source,
an infinitely fine sampling, and, apart from the inevitable photon shot noise, no additional
noise introduced by atmospheric effects or the readout characteristics of the detector.

The phasing in a segmented mirror is done in two steps. First, after the installation
of new segments, initial large \boxheights\ of the order of possibly several wavelengths
have to be reduced to steps of the order of a fraction of a wavelength.
Usually, this only needs to be performed once and can therefore be done
with bright stars. Since they supply a large number of detectable photons,
high accuracies can be obtained during this process. Second, during observations,
one has to be content with small number of photons, but has to correct only comparatively small
\boxheights.
The precision limits or \Cramer\ bounds therefore need to be calculated only for the
case of very small \boxheights\ or \stepheights, that is for $\param \rightarrow 0$.

\subsection{Shack-Hartmann method}
\label{sec:idealSH}
Usually, the Shack-Hartmann method used for phasing measurements consists of placing
a circular lens across the border between two segments. The information on the
\stepheight\ is contained in the diffraction pattern of the lens.
In a one-dimensional setting,
a cylindrical lens is used with its axis being parallel to the border,
and its center ideally exactly on the border. The normal configurations for the
Shack-Hartmann sensor are therefore the single-border and the centred single-border
configuration.
However, to be able to compare the Shack-Hartmann sensor with the other phasing
wavefront sensors, the computation of the \PDF\
in the focal plane and the \EFI,
presented in appendix A1, are done also for the general configuration
shown in figure~\ref{fig:step}, that is for a single lens covering
the full aperture with one out-of-phase segment in an arbitrary location.
The \PDF\ $p(\diffAngle\cond\param)$ and the  \EFI\ $i_{{\rm F}}(\diffAngle\cond\param)$
are both functions of the \stepheight~$\param$, and a
is a dimensionless position variable $\xi$, which is related to a true position
variable $\varDet$ in the focal or detector plane by $\xi = \varDet/f$, where $f$ is the
focal length of the lens.
The general as well as the special configurations will be discussed in this section.

Section~\ref{sec:idealSHFID} discusses the \PDF, the \FID, and the \EFI.
A variation of the parameter $\param$ causes a shift of the maximum
of the diffraction pattern as well as a change of its form.
From both characteristics
partial information about $\param$ can be obtained. Section~\ref{sec:SHIdealShift} shows
that most of the information is contained in the shift of the maximum,
and section~\ref{sec:SHIdealStructure} that a quarter of the full information
is contained in the change of the {\it form}, that is, independent of the horizontal
position of the \PDF.
\begin{figure}
     \centerline{\hbox{
     \psfig{figure=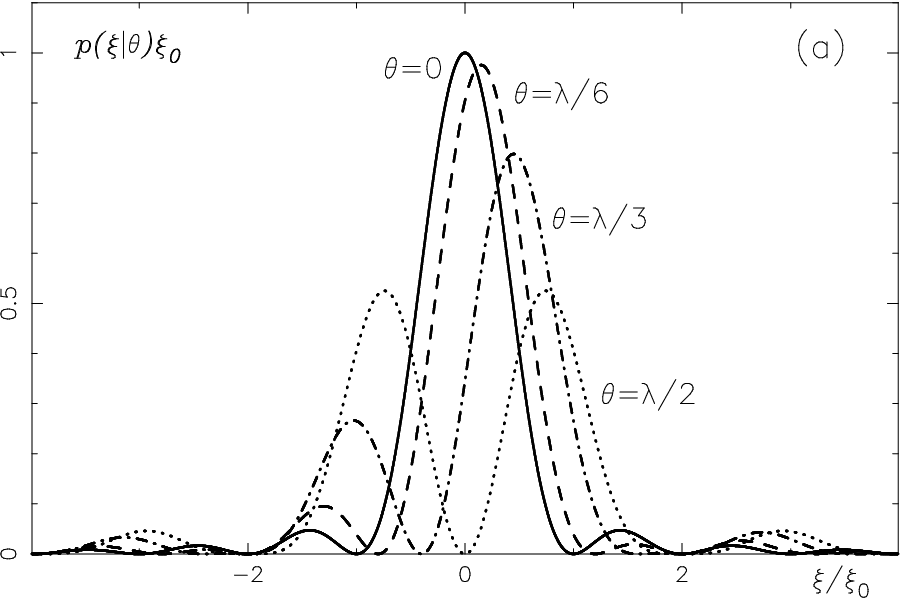,width=85mm}
     \psfig{figure=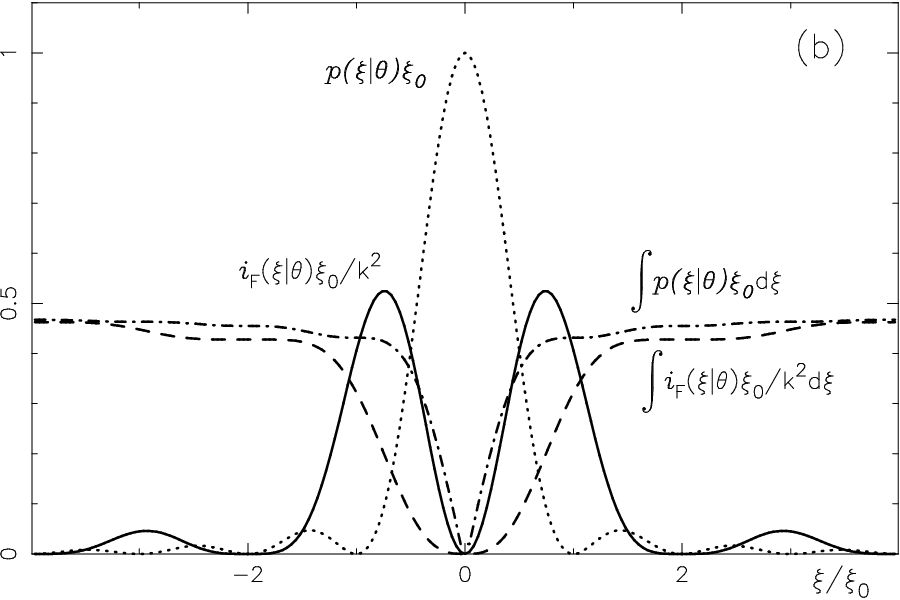,width=85mm}}}
     \caption{\label{fig:SHIdealPdfFid} {\small
      {\bf Shack-Hartmann sensor:} (a) Likelihood functions for four
       \boxheights\ $\param$ as a function of a the normalised angular coordinate
       in the focal plane of the Shack-Hartmann lens.
       (b) Likelihood function and corresponding functions related
       to the Fisher information for $\param = 0$:
       normalised \PDF\ $p(\diffAngle\cond\param)\xio$ (dotted),
       normalised \FID\ $i_{\rm F}(\diffAngle\cond\param)\xio/k^{2}$ (solid),
       integrated \PDF\  and {\it dashed :} integrated \FID\
       divided by $k^{2}$ )dashed-dotted).}}
\end{figure}
\subsubsection{Likelihood function, \FID\, and \EFI\ }
\label{sec:idealSHFID}
Figure~\ref{fig:SHIdealPdfFid}a shows for the centred single-border configuration
the diffraction patterns, or equivalently the \PDFs\ $p(\diffAngle\cond\param)$,
for four different \stepheights~$\param$ ranging from zero to~$\lambda/2$.
The location of the first zero in the diffraction pattern
for $\param=0$ will be denoted by $\xio=\lambda/(2a)$.
For the centred single-border configuration and for arbitrary $\param$ the \FID\
and the \PDF\ are related by
\begin{equation}
  i_{{\rm F}}(\diffAngle\cond\param) = k^{2} \, \p(\diffAngle\cond\param-\lambda/2)
  \quad.
  \label{eq:intensityVersusFisher}
\end{equation}
Figure~\ref{fig:SHIdealPdfFid}b shows that for the case $\param = 0$
the first maxima of the \FID\ on both sides
of the origin are lying just inside the first minima of the \PDF,
represented by the dotted curve. 
The dashed line in
figure~\ref{fig:SHIdealPdfFid}b shows the \FID\ integrated
from~$\diffAngle = 0$ to the~$\diffAngle$ value on the abszissa
and divided by $k^{2}$,
and the dashed-dotted line the corresponding integrated \PDF.
The \PDF\ is more concentrated around the origin than
the corresponding \FID. For the case $\param = \lambda/2$
the \PDF\ has a minimum at $\diffAngle = 0$ and the \FID\ consequently
its maximum at $\diffAngle = 0$.
Therefore, for $\param = \lambda/2$, the \FID\ is more concentrated around
the origin than the corresponding \PDF.

The \EFI\ is obtained by integrating the \FID\ from $\diffAngle = -\infty$ to
$\diffAngle = +\infty$.
For the general configuration
in the limiting case $\param \rightarrow 0$ the integration
of the \FID\ given in equation~(\ref{eq:fidSHSingle}) can be done analytically
and yields
\begin{equation}
  I_{{\rm F,SH}} = k^{2} \, \frac{2}{a} \, {\rm min}(\halfSegWidth,|\centerSegment|)
     \quad.
  \label{eq:fisherInformationSH}
\end{equation}
Since the maximum of ${\rm min}(\halfSegWidth,|\centerSegment|)$ equals $a/2$,
the Fisher information in equation~(\ref{eq:fisherInformationSH}) is always bounded
by the quantum statistical information given in equation~(\ref{eq:efiQMmax}).
For the special case of the single-border configuration this can be seen in
figure~\ref{fig:efiCompareSensors}a where the dashed line shows the linear dependence of
the \EFI\ for the Shack-Hartmann sensor on the segment width $2\halfSegWidth$.

For the general configuration and for a given $\halfSegWidth=0.1a$,
figure~\ref{fig:efiCompareSensors}b shows the \EFI\ for the
Shack-Hartmann sensor
as a function of the location of the segment centre.
Surprisingly the \EFI\ is zero for a segment in the centre
of the aperture.

The \EFI\ for the Shack-Hartmann sensor attains its maximum
for the centred single-border configuration,
that is for $\halfSegWidth=|\centerSegment|=a/2$:
\begin{equation}
  I_{{\rm F,SH,max}} = k^{2}
  \label{eq:fisherInformationSHMaximum}
    \quad ,
\end{equation}
and is identical to the quantum statistical \EFI.
Therefore, if all of the relevant information about the \boxheight\ in the diffraction
pattern could be extracted with an efficient estimator, one could reach the maximum
possible precision, that is the fundamental bound (\ref{eq:efiQMmax})
for the estimate of the \stepheight~$\param$.

The formulae given above have been derived for one lens covering the full aperture
$[-a,+a]$. However, in a phasing wavefront sensor, a single lenslet will only cover
a subaperture across a border. For these lenslets
the configuration will ideally be the centred single-border configuration.
It can readily be verified that due to the linear dependence
of the \EFI\ on the segment width, the information $I_{{\rm F}}$ obtained
with a single lens across the full aperture is identical to the
information obtained with lenslets across the borders and centred on the borders,
provided that
the lenslets cover the full segment width and the segment centre
is not too close to the centre of the aperture.
Since the other sensors work over the full aperture, the \EFI\ for
the Shack-Hartmann method can therefore be compared with the other methods by using
the results for a single lens covering the full aperture.
\subsubsection{Information contained in the shift of the maximum of the \PDF}
\label{sec:SHIdealShift}
Figure~\ref{fig:SHIdealPdfFid}b, applying to the centred single-border configuration,
shows that in the most interesting case
of~$\param \ll \lambda$ nearly 50\% of the \EFI\ is
contained in the interval between the first
zeroes of the \PDF\ on both sides of the origin. The maximum of the \PDF\ is
well defined, even in the case of broadband measurements or in the presence of
atmospheric disturbances as shown in sections~\ref{sec:broadband}
and \ref{sec:atmosphere}. Furthermore,
the position of the maximum depends on the \stepheight~$\param$.
In addition, the position of the maximum is not too sensitive
to small misalignments between
the centre of the lens and the border between segments.
For small \stepheights~$\param$ and small deviations from the centred single-border
configuration, described by the difference $\centerSegment-a/2$,
that is for $k\param \ll
1$ and~$k(\centerSegment - a/2) \ll 1$, one obtains for the
location~$\diffAngle_{\rm max}(\param)$ of the maximum of the \PDF\
\begin{equation}
  \diffAngle_{\rm max}(\param) = -\frac{3}{4} \; \frac{\param}{a} \;
         (1 - 4\frac{(\centerSegment-a/2)^{2}}{a^{2}})
     \quad.
     \label{eq:maxOmega}
\end{equation}
This equation shows that the location of the maximum is only
weakly affected by the misalignment~$\centerSegment - a/2$ provided
that $\centerSegment - a/2 \ll a$.
A similar expression for $\diffAngle_{\rm max}(\param)$
without the effect of the decentring has been given in \cite{Chanan2000}
for circular lenslets.

For all these reasons,
it is of interest to estimate the \stepheight~$\param$ only from
the core of the \PDF\, which is defined here as the interval where the \PDF\ exceeds
20\% of its maximum value at the centre of the core.
Figure~\ref{fig:SHIdealPdfFid}b shows that for the centred single-border configuration
the integral of the \FID\ over that interval is approximately $0.3I_{{\rm F,SH,max}}$,
that is approximately 30\% of the maximum information can be extracted from the core.

The central core can quite accurately be fitted by a Gaussian
\begin{equation}
  p_{{\rm gauss}}(\diffAngle\cond\param) = \frac{0.6 ka}{\sqrt{\pi}}
           \, e^{-[0.6 ka(\diffAngle-3\param/(4a))]^{2}}
  \quad.
  \label{eq:fittedGauss}
\end{equation}
Its \EFI\ is given by
\begin{equation}
  I_{\rm F,gauss} = 0.4\, k^{2} \approx 0.4\, I_{\rm F,SH,max}
  \quad.
  \label{eq:fisherGauss}
\end{equation}
This is only marginally larger than the information content $0.3I_{{\rm F,SH,max}}$
in the core in the mentioned above.
The reason for the small difference is probably that the integration of the Gaussian
probability function extends from $-\infty$ to $+\infty$, whereas the one of the
diffraction pattern only over the central core.

The shift~$\diffAngle_{\rm mean}$ of the mean position, i.e.\ the
centre of gravity of the \PDF, is not identical to the shift of the
central maximum of the diffraction pattern. $\diffAngle_{\rm
mean}$ should be smaller than~$\diffAngle_{\rm max}$ since the shift
of $\diffAngle_{\rm max}$ to one side is accompanied by an increase
of the first peak on the opposite side.
For the centred single lens configuration one obtains from
equation~(\ref{eq:intensitySH}) for the centre of gravity
\begin{equation}
  \diffAngle_{\rm mean} = -\frac{1}{2ka} \, \sin(k\param)
  \quad.
            \label{eq:shiftCOG2}
\end{equation}
For~$k\param \ll 1$ this becomes
\begin{equation}
  \diffAngle_{\rm mean} = -\frac{\param}{2a}
  \quad.
            \label{eq:shiftCOGSmall}
\end{equation}
A comparison with equation~(\ref{eq:maxOmega}) shows that
this is indeed smaller than the shift~$\diffAngle_{\rm max}$ of the
maximum by a factor of~$1.5$. It is equal to half of the average
tilt of the wavefront $\varphi(\varPupil)$ over the aperture in the pupil.

\subsubsection{Information contained in the form of the \PDF}
\label{sec:SHIdealStructure}
In some applications a reference pattern generated by the Shack-Hartmann lenslet array
may not be available. Furthermore,
an adaptive optics system may partially correct a phase step, in the simplest
case by introducing a constant tilt over a subaperture covered by a lenslet.
Such a tilt, the amount of which may not be known, will lead to a shift
of the diffraction pattern.
Under such circumstances
no information can be obtained from the position of the
diffraction pattern, and the
parameter~$\param$ can only be estimated from the changes in the
form of the \PDF.

\begin{figure}[h]
  \centerline{\hbox{
   \psfig{figure=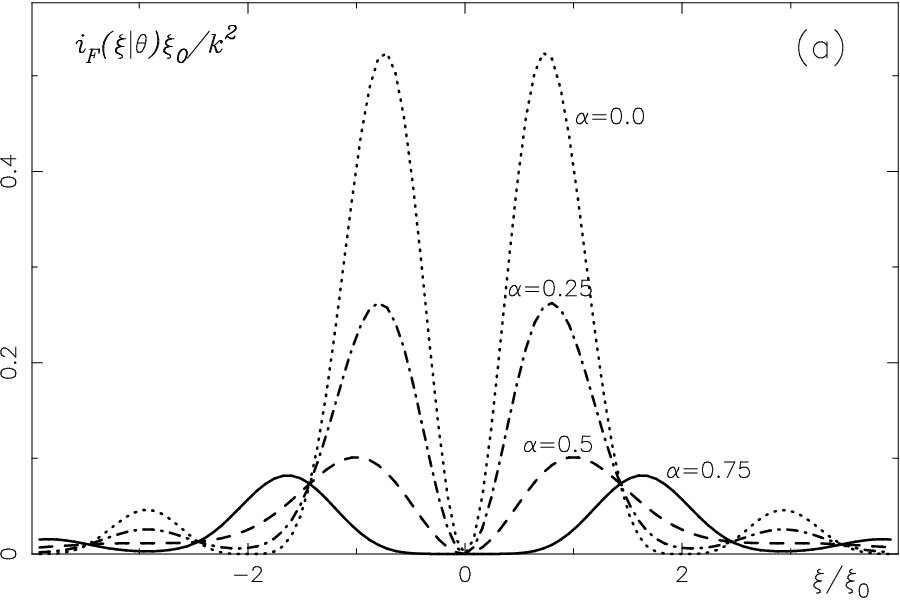,width=85mm}
   \psfig{figure=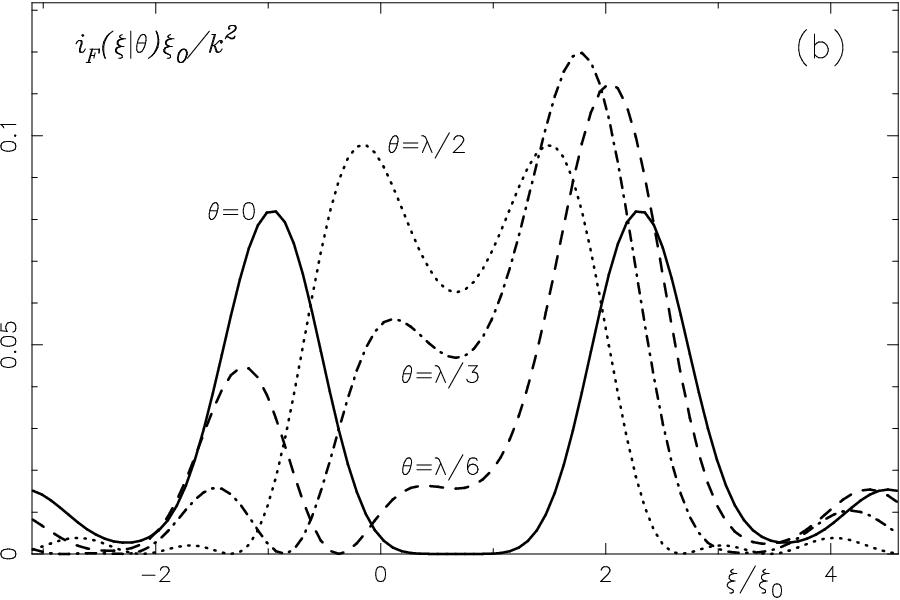,width=85mm}}}
   \caption{\label{fig:fisherCompensatedshift} {\small
       {\bf Shack-Hartmann sensor:} 
      (a) Normalised \FID\ $i_{\rm F}(\diffAngle\cond\param) \xio/k^{2}$
            compensated by various shifts of~$\alpha \param/a$
           for the limit~$\param \rightarrow 0$. The values for $\alpha$ are
           given in the figure.
     (b) Normalised  \FID\ in the form of the \PDF\ 
           for four \stepheights~$\param$.}}
 \end{figure}

The \EFI\ contained in the form, from now on called the
form-only \EFI, can be calculated as follows.
The likelihood function for a given \stepheight~$\param$
can be shifted such that its maximum is closer to the origin.
The \EFI\ can then be calculated as
a function of this shift, with the information contained in the form
being defined as the minimum of the \EFI\ as a function of the shift.
For the centred single-border configuration a \PDF~$\p_{\rm s}$
shifted by~$\alpha \param/a$ is described by
\begin{equation}
  \p_{\rm s}(\diffAngle\cond\param)  = \frac{1}{\xio} \,
               \frac{1}{(\pi\diffAngle/\xio - \alpha2\pi\param/\lambda)^{2}}
               \frac{1}{(\diffAngle -\alpha \param/a)^{2}}\;
                \left[ \sin(\pi\diffAngle/\xio - \alpha2\pi\param/\lambda
                             + 2\pi\param/\lambda)
                            - \sin(2\pi\param/\lambda) \right]^{2}
    \quad.
    \label{eq:intensity2Compensated}
\end{equation}
Figure~\ref{fig:fisherCompensatedshift}a shows for a few values of~$\alpha$
the \FIDs\ for $\param=0$.
For arbitrary values of~$\param$ the minimum \EFI\ is
analytically obtained for~$\alpha = 3/4$. 
According to equation~(\ref{eq:maxOmega}), for $\param \ll \lambda$
such a shift is equivalent to moving the maximum of the
\PDF\ back to~$\diffAngle=0$.
Figure~\ref{fig:fisherCompensatedshift}b shows the form-only \FIDs\
for four \stepheights\ ranging from $\param=0$ to $\param=\lambda/2$.
A comparison with figure~\ref{fig:SHIdealPdfFid}b shows that for $\param=0$
the peaks of the \FIDs\ are much lower than in the case of the full information
and are further from the origin.

For all values of $\param$ the minimum \EFI\ in the centred single-border configuration
is given by
\begin{equation}
  I_{{\rm F,SH,form}} = k^{2} / 4
  \quad.
           \label{eq:fisherInfComp}
\end{equation}
Therefore, if an estimate of the \stepheight\ $\param$ can only be based
on the form of the \PDF, one
needs four times the number of photons to reach the same precision
as is obtainable from the full information, which includes the information
from the position.
\subsection{Mach-Zehnder interferometer}
\begin{figure}[h]
  \centerline{\hbox{
   \psfig{figure=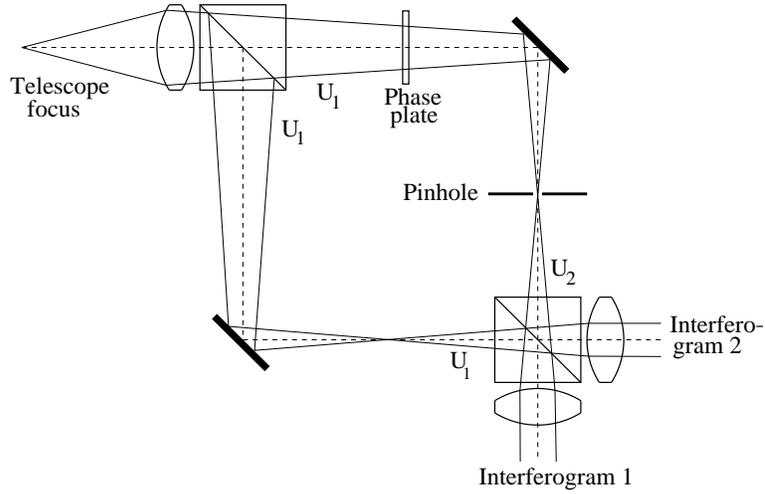,width=100mm}}}
   \caption{\label{fig:principleMZ}
     {\small {\bf Mach-Zehnder sensor:} Principle of a Mach-Zehnder interferometer
      as used for the detection of \boxheights.}}
 \end{figure}

\begin{figure}[h]
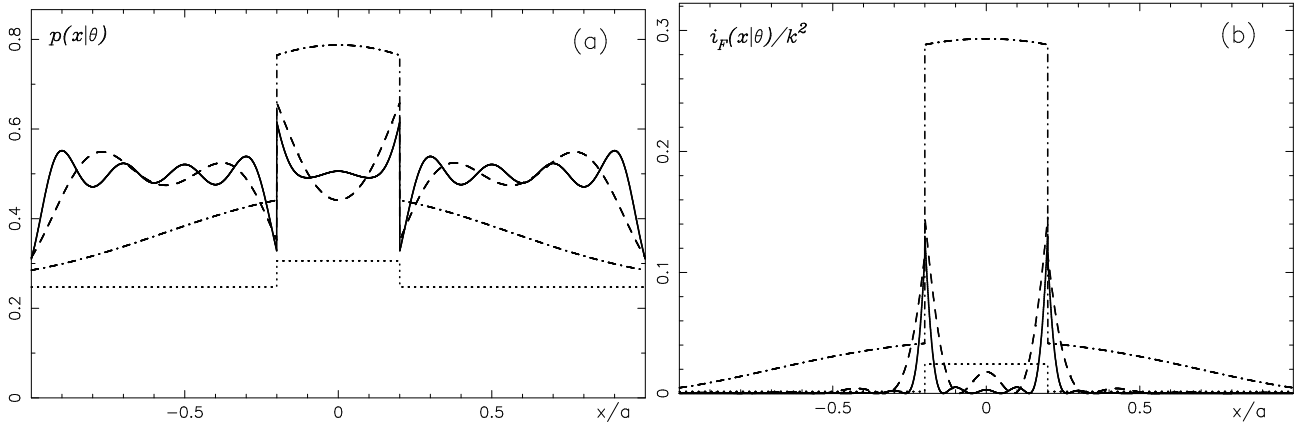

  \centerline{\hbox{
   \psfig{figure=pdfMZSharp.ps,width=85mm}
   \psfig{figure=fidMZSharp.ps,width=85mm}}}
   \caption{\label{fig:pdfFidMZSharp}
     {\small {\bf Mach-Zehnder sensor:} 
      (a) Likelihood functions generated by a Mach-Zehnder interferometer
        with sharp edge pinholes with radii of 0.1 (dotted),
        0.8 (dashed-dotted),
        4 (dashed) and 10 times $\lambda/(2a)$ (solid), for a \boxheight\
        $\param = \lambda/10$.
      (b) Corresponding normalised \FIDs\ $i_{\rm F} / k^{2}$ for $\param=0$.}}
 \end{figure}

\begin{figure}[h]
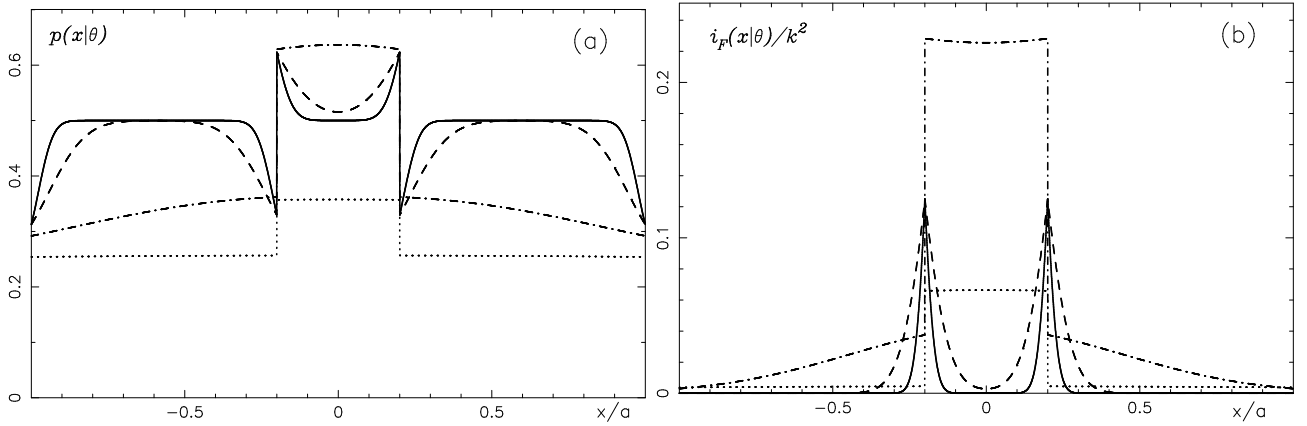

  \centerline{\hbox{
   \psfig{figure=pdfMZGauss.ps,width=85mm}
   \psfig{figure=fidMZGauss.ps,width=85mm}}}
   \caption{\label{fig:pdfFidMZGauss}
     {\small {\bf Mach-Zehnder sensor:}
         (a) Likelihood functions generated by a Mach-Zehnder interferometer
           with Gaussian pinholes with RMS values of 0.1 (dotted),
           0.4 (dashed-dotted),
           2 (dashed) and 5 times $\lambda/(2a)$ (solid),
           for a \boxheight\ $\param = \lambda/10$.
        (b) Corresponding normalized \FIDs\ $i_{\rm F}/k^{2}$ for $\param=0$.}}
 \end{figure}

The principle of the Mach-Zehnder interferometer as used for the
detection of \boxheights\ \cite{Yaitskova2005} is shown
in figure~\ref{fig:principleMZ}.
The incoming light is focussed and split into two beams.
One of them, which serves as the reference beam,
contains a spatial filter in the focal plane and a phase plate
introducing a phase shift $\phi$.
The light from the two beams is then
recombined behind a second beam splitter, collimated
and registered on two detectors in planes which are conjugated
to the entrance pupil.
The spatial filter is either a pinhole with a sharp edge or with a
Gaussian transmission function with 100\% transmission at the centre.
If $\xi$ is an angular coordinate in the focal plane, the transmission functions
of the filters are given by
\begin{eqnarray}
   t(\xi) & = &
      \left\{ \begin{array}{ll}
              \left\{
                 \begin{array}{lll}
                  1 & {\rm for} & |\xi| \le \zeta\\
                  0 & {\rm for} & |\xi| > \zeta
                 \end{array}
              \right\} &  {\rm Sharp-edge}  \vspace*{2mm} \\
  \hspace{5mm} \exp(-\xi^{2}/(2\zeta^{2})) & {\rm Gaussian} 
  \end{array} \right\}
     \quad ,
     \label{eq:shapePinhole}
\end{eqnarray}
where $\zeta$, also in angular coordinates,
is either half of the diameter of the sharp pinhole or
the RMS of the Gaussian transmission function.
The computation of the \PDF\ and the \FID\ is outlined in appendix A2.
For a sharp-edge pinhole and $\param=\lambda/10$,
figure~\ref{fig:pdfFidMZSharp}a shows the \PDFs\ for
four different half-diameters $\zeta$ of the sharp-edge pinhole,
expressed as fractions or multiples of $\xio=\lambda/(2a)$,
and figure~\ref{fig:pdfFidMZSharp}b the corresponding \FIDs.
Similar curves for a Gaussian pinhole
are shown in figures~\ref{fig:pdfFidMZGauss}a and \ref{fig:pdfFidMZGauss}b.

Obviously, the \EFI\ is maximised for some intermediate values of $\zeta$.
On the one hand, for very small pinhole widths the form of the likelihood functions
resembles closely the
form of the phase in the wavefront after the reflection by the mirrors.
However, the Fisher information is small, since only a small fraction of the light
entering the second beam interferes with the light from the reference beam. In the limit
$\zeta \rightarrow 0$ one has $I_{{\rm F,MZ}} \propto \zeta^{2}$.
On the other hand, for large pinholes the form in the interference pattern and the \FID\
are concentrated around the borders between the segments. In the limit of very large pinhole sizes
the two beams are, apart from a constant phase shift, identical and the information content
goes to zero as $1/\zeta$ for $\zeta \rightarrow \infty$. The maximum Fisher
information must therefore be obtained for an intermediate size of the pinhole.
For a segment half-width of $w=0.2a$ this maximum is obtained for $\zeta \approx 0.8\xio$
in the case of a sharp pinhole and for $\zeta \approx 0.4\xio$ in the case of
a Gaussian pinhole, indicated by the dashed-dotted lines in the
figures~\ref{fig:pdfFidMZSharp} and ~\ref{fig:pdfFidMZGauss}, respectively.

\begin{figure}[h]
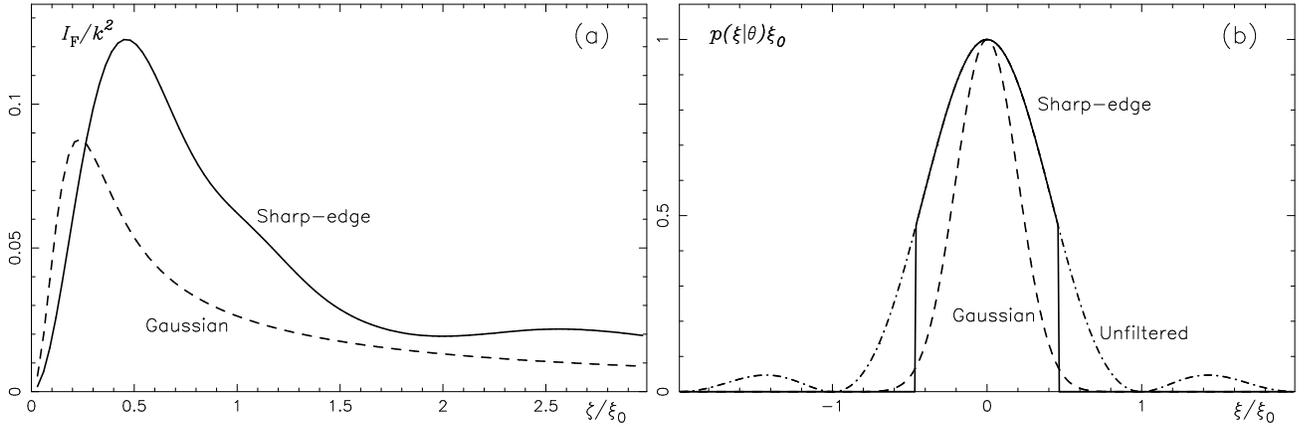

  \centerline{\hbox{
   \psfig{figure=efiMZ.ps,width=85mm}
   \psfig{figure=filters.ps,width=85mm}}}
   \caption{\label{fig:efiMZ}
     {\small {\bf Mach-Zehnder sensor:}
     (a) Normalised Fisher information $I_{{\rm F,MZ}}/k^{2}$ as a function
          of the normalised pinhole radius $\zeta/\xio$ for pinholes with a sharp edge
         (solid) and with a Gaussian transmission function ({\it dashed}) for the centred
          single-border configuration.
     (b) Intensities filtered with a sharp-edge pinhole (solid), and with a Gaussian
         pinhole (dashed) with widths which maximise the Fisher information.
         The dashed-dotted curve represents the unfiltered diffraction pattern
         for a \stepheight\ $\param=0$.}}
 \end{figure}

For the centred single-border configuration the maxima can be obtained from
figure~\ref{fig:efiMZ}a which shows the normalised sum $I_{{\rm F}}/k^{2}$ of the
Fisher information in the two beams as a function of the ratio of $\zeta$ to $\xio$ .
In this case the maximum of the Fisher information is obtained for
$\zeta \approx 0.465 \xio$ in the case of a sharp-edge pinhole and
$\zeta \approx 0.235 \xio$ in the case of Gaussian pinhole.
Figure~\ref{fig:efiMZ}b shows that these sizes of the pinholes maximising the \EFI\
are for both types
of pinholes comparable to the size of the core of the image in the focal plane.
However, even under ideal conditions the maximum Fisher information
is only approximately 12\% of the quantum statistical \EFI\ in the case of the
sharp-edge pinhole and 9\% in the case of the Gaussian pinhole.

For the single-border configuration and a width of the pinhole which maximizes the \EFI\
the dotted line in figure~\ref{fig:efiCompareSensors}a shows that
$I_{{\rm F,MZ}}(\param)$ depends quadratically on the segment width, and
figure~\ref{fig:efiCompareSensors}b shows that for the general configuration
the \EFI\ is slighly larger for segments closer to the centre of the aperture.

\begin{figure}[h]
  \centerline{\hbox{
   \psfig{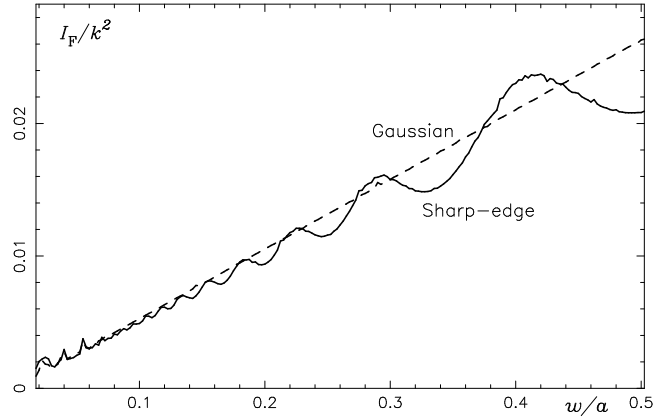}}}
   \caption{\label{fig:efiVarWMZ}
     {\small {\bf Mach-Zehnder sensor:}
          Normalised Fisher information $I_{{\rm F,MZ}}/k^{2}$ as a function
          of half of the segment width $w$ in a single-border configuration
          with pinhole sizes of $\lambda/w$
          for the sharp edge pinhole (solid), and $\lambda/(2w)$ for the Gaussian
          pinhole (dashed).}}
 \end{figure}

However, if the Mach-Zehnder method is applied in a telescope with several segments, the signals
from neighbouring borders have to be well separated. Quite arbitrarily it will be defined that
this condition is fulfilled, if the distances between the border and,
in the case of the sharp-edge pinhole, the first zero, or, in the case of the Gaussian pinhole, 
the $x$-coordinate where the signal has dropped
to 10\% of the value at the border,
are both approximately equal to $w/3$, as shown by the solid lines in the
figures~\ref{fig:pdfFidMZSharp} and \ref{fig:pdfFidMZGauss}.
This requires widths of $\zeta = \lambda/w$
for the sharp-edge pinhole and $\zeta = \lambda/(2w)$ for the Gaussian pinhole.
Figure~\ref{fig:efiVarWMZ} shows that with these choices of the pinhole widths
the \EFI\ for the single-border configuration is now proportional
to the segment width $w$.
However, for such widths of the pinholes the \EFI\ is much lower than for the widths
maximising the \EFI.

For well separated signals the \FIDs\ in the neighbourhood of a border
are independent of the segment position.
As long as the signals from both borders in a two-border configuration are well within
the aperture, the \EFI\ is therefore also independent of the segment location.
\subsection{Foucault knife-edge test}
\Boxheights\ can also be measured by a Foucault test. The incoming beam is focussed,
a knife-edge is placed in the centre of the focal plane, and finally the beam is collimated
and detected in a plane conjugated to the pupil. By replacing the knife-edge with
a glass prism both halves of the beam can be used for the analysis
\cite{Esposito2001}\cite{Esposito2003}\cite{Pinna2004}.
The computation of the \PDF\ and the \FID\ is outlined in appendix A3.
Figure~\ref{fig:pdfFidFC}a shows the \PDFs\ for three locations of the segment centre
with a segment half-width of $\halfSegWidth = 0.2a$ and for $\param /\lambda=0.1$,
and figure~\ref{fig:pdfFidFC}b the \FIDs\ for $\param = 0$.
The widths of the peaks in the \FIDs\ in the neighbourhood of the borders
are approximately equal to the widths of the peaks
in the \PDFs. The ratio of the width of the \FID\ in the neighbourhood of a border
to the width $2\halfSegWidth$ of the
segment is independent of the other parameters and approximately equal to 0.2.
The information from adjacent borders can therefore always be well separated.

 \begin{figure}[h]
  \centerline{\hbox{
    \psfig{figure=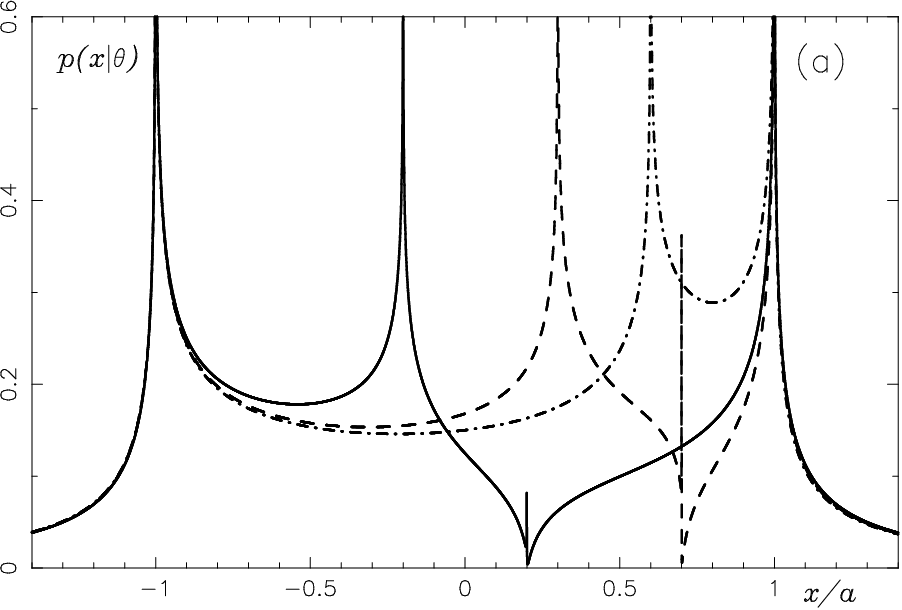,width=85mm}
    \psfig{figure=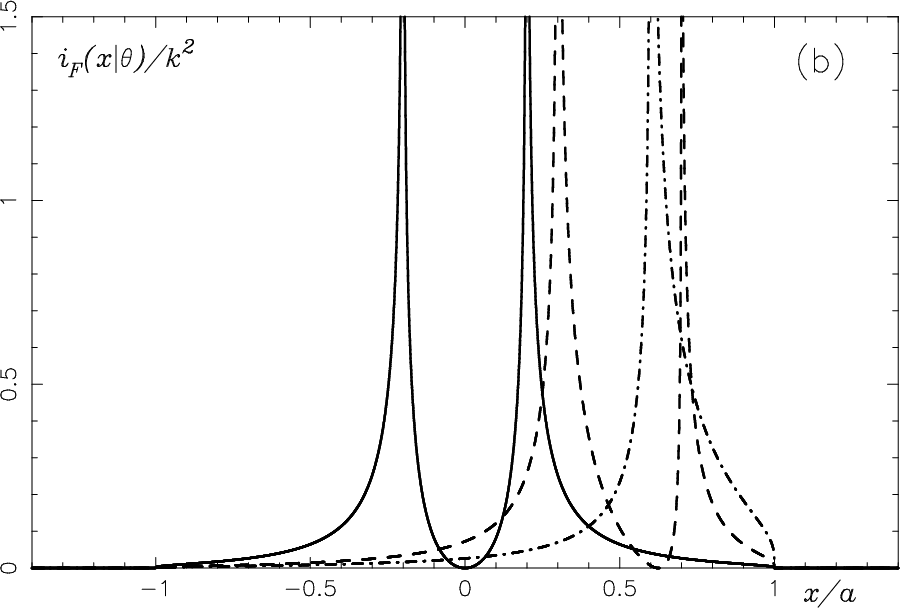,width=85mm}}}
    \caption{\label{fig:pdfFidFC} {\small
     {\bf Foucault sensor:} 
      (a) Likelihood function for a \boxheight\ $\param=\lambda/10$.
      (b) Normalised \FID\ for a \boxheight\ $\param = 0$ and $\halfSegWidth = 0.2a$,
         and, for both plots,
         three different positions $c=0$ (solid), $c=0.5a$ (dashed)
         and $c=0.8a$ (dashed-dotted) of the segment centre.}}
  \end{figure}
A numerical integration shows that for the centred single-border configuration
the \EFI\ in one of the exit pupil equals $k^{2}/4$.
This may intuitively be expected, since only half of the light enters
one of the two exit pupils and only half of the image in the focal plane is used
to retrieve information. Adding up the information in both channels gives
$I_{{\rm F,FC}}=k^{2}/2$. The Foucault sensor with a glass prism therefore
extracts potentially half of the quantum statistical information.

The dashed-dotted line in figure~\ref{fig:efiCompareSensors}a
shows the \EFI\ as a function of the
segment width in the single-border configuration. The dependence is quadratic, similar
to the case of the quantum statistical information. For all segment widths the \EFI\ of the
Foucault sensor equals half of the quantum statistical information.
For the general configuration figure~\ref{fig:efiCompareSensors}b shows that the \EFI\ is
effectively independent of the segment location. Figure~\ref{fig:pdfFidFC}b shows
therefore that the \FID\ is larger in the neighbourhood of the single-border
in the single-border configuration
than in the neighbourhood of the two borders in the general configuration.
\subsection{Curvature sensor}
\label{sec:curvatureSensor}
\begin{figure}[h]
  \centerline{\hbox{
    \psfig{figure=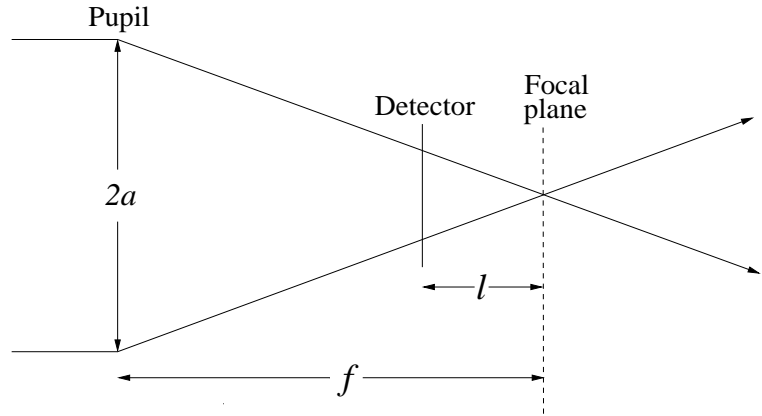,width=100mm}}}
    \caption{\label{fig:curvature} {\small
         {\bf Curvature sensor:} Principle of the curvature wavefront sensor with
          the detector plane being out-of-focus by a distance $l$ with respect
          to the focal plane of the telescope.}}
  \end{figure}

\begin{figure}[h]
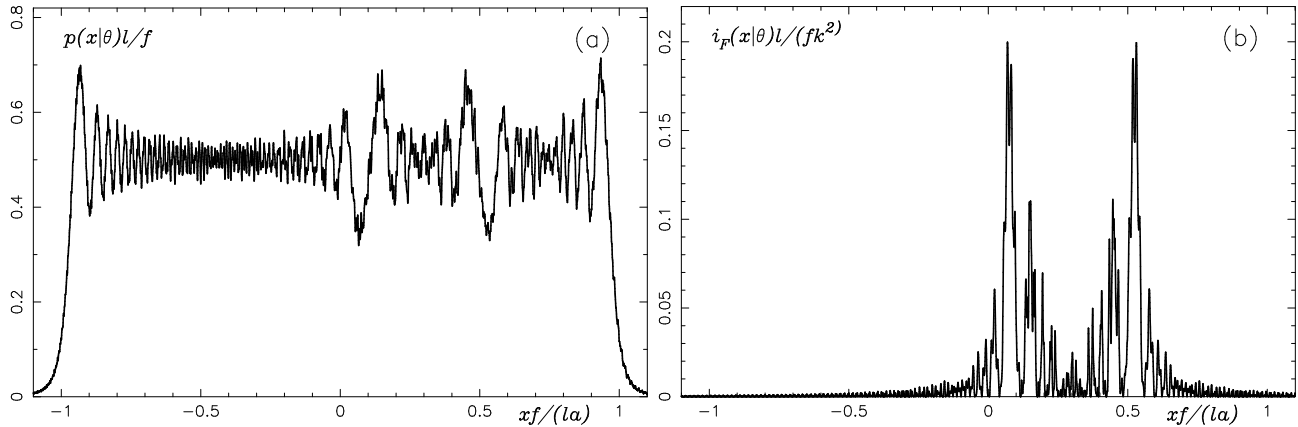

  \centerline{\hbox{
    \psfig{figure=pdfCV.ps,width=85mm}
    \psfig{figure=fidCV.ps,width=85mm}}}
    \caption{\label{fig:CVPdfFid} {\small
           {\bf Curvature sensor:} 
          (a) Normalised \PDF\ for a defocussing ratio $l/f = 0.002$
              and $\param=\lambda/10$.
          (b) Normalised \FID\ for $l/f = 0.002$, both as functions of the
              normalised detector coordinate $xf/(la)$.
          For both plots : $a=1m$, $f=20m$, $c=0.3a$, and $w=0.2a$.}}
  \end{figure}

In a curvature sensor the detector is placed at a distance $l$ in front
or behind the focal plane, as shown in figure~\ref{fig:curvature}.
Since the diffraction pattern
is spread over an interval of approximately $2af/l$ and the intensities are therefore
approximately proportional to $l/f$, the detector coordinates in the figures are normalised
by a multiplication with $f/(la)$ and the \PDF\ and the \FID\ are normalised
by a multiplication with $l/f$.
Except for detector positions very close to the
aperture, the diffraction pattern can be calculated with the Fresnel approximation,
which is outlined in appendix A4.
In the limit $l \rightarrow 0$ the expressions for the likelihood functions and
for the \FIDs\ converge to the respective expressions for the Shack-Hartmann sensor
derived in appendix A1.
However, the curvature method requires the signals from
the two segment borders to be clearly separated. In other words, the \FIDs\
corresponding to the two borders should not overlap.
For $\halfSegWidth = 0.2a$ such a separation in the \PDF\ as well as in the
\FID\ seems to be established for $l/f \ge 0.002$, as shown in the
figures~\ref{fig:CVPdfFid}a and \ref{fig:CVPdfFid}b.
 \begin{figure}[h]
  \centerline{\hbox{
    \psfig{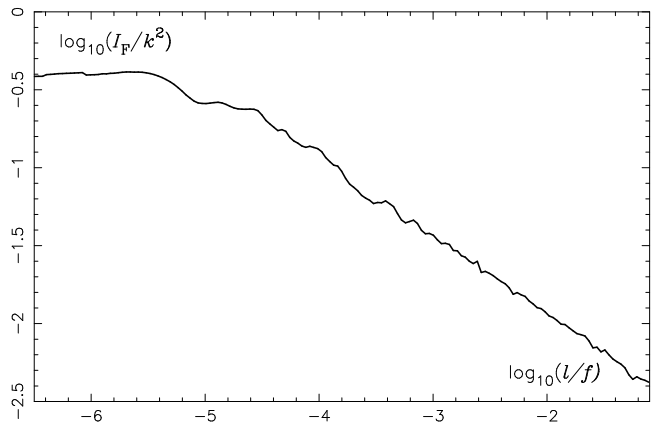}}}
    \caption{\label{fig:CVDefocus} {\small
           {\bf Curvature sensor:} 
              Normalised Fisher information as a function of the
              defocussing ratio $l/f$,
              and $a=1m$, $f=20m$, $c=0.3a$, and $w=0.2a$.}}
  \end{figure}

Figure~\ref{fig:CVDefocus} shows in a log-log-plot the \EFI\ as a function of the
ratio $l/f$, and for the same parameters as in figure~\ref{fig:CVPdfFid}.
The regime where $I_{{\rm F,CV}}$ is approximately constant is the Fraunhofer regime.
Let the pure curvature regime be defined as the regime where the signals
from two adjacent borders are well separated. The ratio $l/f=0.002$,
which is the lower limit for which a separation is guaranteed,
lies at the beginning of the regime with a constant slope in the log-log-plot,
where $I_{{\rm F,CV}} \propto \sqrt{f/l}$ and which extends at least up to
$l/f\approx 0.1$. With a choice of $l/f>0.002$ one is well outside the caustic,
that is, the correlation of the subapertures is identical with that in the pupil.

A precise definition of the signal width which guarantees
a separation of signals from adjacent borders is somewhat arbitrary.
Requiring the distance of the first zero of the
oscillating pattern 
to be less than a fraction $1/z$ of the segment half-width $\halfSegWidth$
leads to the condition
\begin{equation}
 \frac{l}{f} > \left( 1+\frac{\halfSegWidth^{2}}{0.8z^{2}\lambda f}\right)^{-1}
     \quad .
   \label{eq:condNonOverlap}
\end{equation}
If, for different widths $2\halfSegWidth$ of the segment, the parameter $l/f$ is always chosen
such that the condition~(\ref{eq:condNonOverlap}) is fulfilled for $z=10$,
the dependence of the \EFI\
on the segment width is linear, as in the case of the Shack-Hartmann sensor, that is
in the Fraunhofer limit. However, for the chosen non-overlap condition~(\ref{eq:condNonOverlap})
the values for the \EFI\ are approximately 30 times lower than in the Fraunhofer limit.

$I_{{\rm F,CV}}(\param)$ is effectively independent of the segment location.
This is similar to the case of the Shack-Hartmann sensor, except that the central dip
in the curve for the Shack-Hartmann sensor in figure~\ref{fig:efiCompareSensors}
is not present in the case of the curvature sensor with sufficient defocussing.
\section{Comparison between the four sensors}
 \begin{figure}[h]
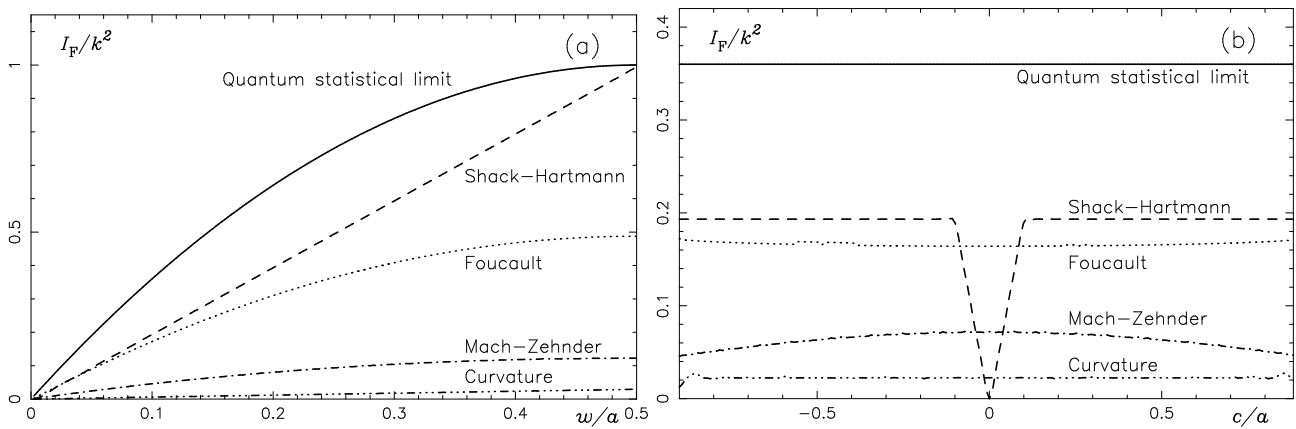

  \centerline{\hbox{
    \psfig{figure=compareWidth.ps,width=85mm}
    \psfig{figure=comparePosition.ps,width=85mm}}}
    \caption{\label{fig:efiCompareSensors} {\small
           {\bf All sensors:} 
      Normalised Fisher information $I_{{\rm F}}/k^{2}$ for
      the single-border configuration as a function of
      (a) the segment width $2\halfSegWidth$ and
      (b) of the segment location for $\halfSegWidth = 0.1a$.
      The curve for the Mach-Zehnder was obtained with
      a sharp-edge pinhole with $\zeta=0.465\xio$. The curves for the curvature sensor
      are calculated for a defocussing which guarantees that the half width 
      of the segment is approximately ten times larger than the distance from
      the coordinate corresponding to a border to the first zero of the oscillating
      pattern.}}
  \end{figure}
The \Cramer\ bounds for the four types of sensors have been discussed in
detail in the preceeding sections. This section provides a comparison
of the relative precision limits which can, in principle, be achieved with these
sensors for the detection of \boxheights. 
Figure~\ref{fig:efiCompareSensors}a shows that for the single-border configuration
under ideal conditions the performance, defined by the
\Cramer\ bound, of the Shack-Hartmann sensor is superior to the performance of the
Foucault sensor, which in turn outperforms the Mach-Zehnder sensor
for all segment widths.
For small segment widths the \EFI\ of the Shack-Hartmann and the Foucault sensors are
effectively identical.
However, except for the Shack-Hartmann sensor in the centred single-border
configuration with $w=a$, none of the sensors attains
the quantum statistical limit.

For the Shack-Hartmann sensor the \EFI\ depends linearly on the segment widths.
The dependence is quadratic for the quantum statistical limit,
the Foucault sensor, and the Mach-Zehnder sensor with pinholes
diameters maximising the \EFI.

For a Mach-Zehnder sensor with large pinhole diameters, which guarantee that
the signals from adjacent borders are well separated, the \EFI\ is a linear function of the
segment width. The maximum value which is attained for the centred single-border configuration
is approximately 0.03 if the signals from adjacent borders are just separated, and decreases
with a better separation of the signals.
Also the \EFI\ of the curvature sensor depends, like the Shack-Hartmann sensor as
its limiting case for small amounts of defocussing, linearly on the segment width.
However, the slope depends strongly on the amount of defocussing, and the curve in
figure~\ref{fig:efiCompareSensors} has been calculated for a defocussing which guarantees
that the half width of the segment is approximately ten times larger than the distance from
the coordinate corresponding to a border to the first zero of the oscillating
pattern. See section~\ref{sec:curvatureSensor} for further details.

Figure~\ref{fig:efiCompareSensors}b shows that for all sensors,
except for the Mach-Zehnder device,
the \EFI\ is effectively independent of the segment location within the pupil.
However, for large pinhole diameters this also applies to the Mach-Zehnder sensor.
The dip in the curve for the
Shack-Hartmann sensor is irrelevant: when such a sensor is used in a
telescope the subapertures would always be defined by smaller lenslets covering only
the neighbourhood around a segment border, and the configuration would then be
the single-border configuration.
\section{Effects of noise and other parameters}
\label{sec:noiseEffects}
The effects of broadband illumination, atmospheric disturbances, detector readout noise,
and discrete sampling of the \PDF\ will only be discussed for the Shack-Hartmann sensor
in the centred single-border configuration. However, the methods used in this section
can readily be applied to all other types of sensors and configurations.
\subsection{Effects of broadband measurements}
\label{sec:broadband}
 \begin{figure}[h]
  \centerline{\hbox{
    \psfig{figure=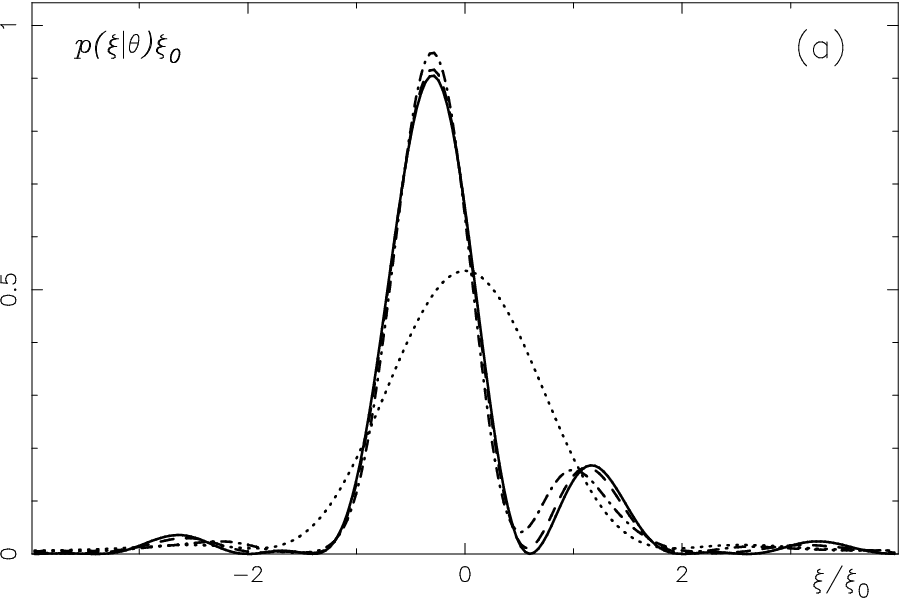,width=85mm}
    \psfig{figure=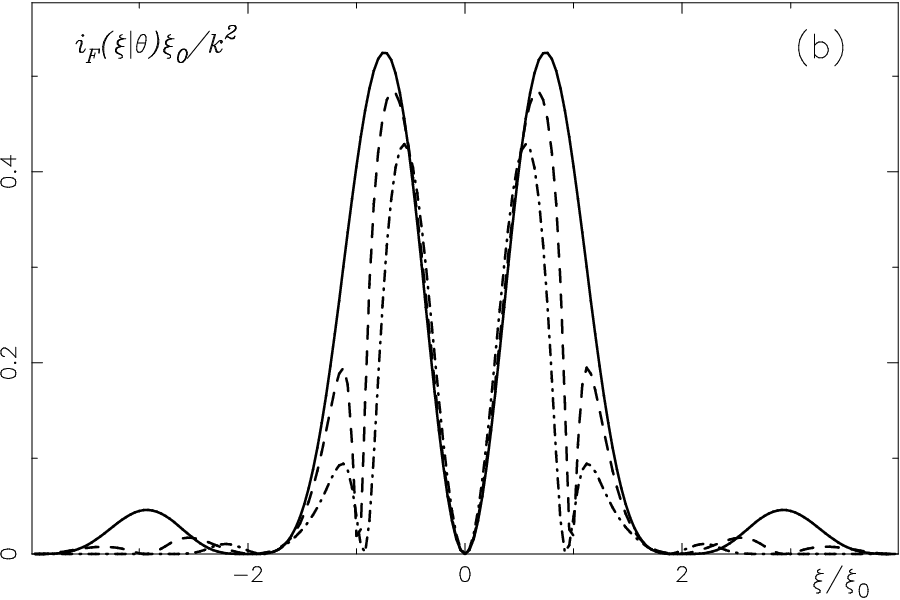,width=85mm}}}
    \caption{\label{fig:pdfFidBroadband} {\small
          {\bf Shack-Hartmann sensor:}
          (a) Likelihood functions for a \stepheight\ $\param=\lambda/5$,
          (b) normalised \FIDs\
             for $\param = 0$, integrated over different bandpass ratios
               $\Delta \lambda / \lambda_{{\rm c}}$:
               Monochromatic (solid),
               $\Delta \lambda / \lambda_{{\rm c}} = 0.25$ (dashed),
               $\Delta \lambda / \lambda_{{\rm c}} = 0.5$ (dashed-dotted).
              The dotted curve in (a) is the limit
             to which the \PDFs\ converge for very large \stepheights,
             that is ~$k\param \gg 1$, and a bandpass ratio of
             $\Delta \lambda / \lambda_{{\rm c}} = 0.5$.}}
  \end{figure}
The number of available photons can be increased with
broadband measurements, that is by using a larger bandpass of the light.
The corresponding gain in the information is, however, partially
compensated by a loss of information due to a loss of form in the \PDF.
As shown below, the \Cramer\ bound decreases with an increase of the bandpass
strongly initially small bandpasses, but only slowly at large bandpasses.

The \PDF\ for a broadband measurement is obtained by integrating the monochromatic
\PDF\ over the wavenumber $k$ around a central wavenumber
$k_{{\rm c}}$ with a suitable prior
probability distribution or spectral density~$p_{{\rm k}}(k)$ as a weight function.
In the following calculations $p_{{\rm k}}(k)={\rm const}$ is assumed for any bandpass.
Figure~\ref{fig:pdfFidBroadband}a shows the \PDF\
for three bandpasses and a \stepheight\ $\param = \lambda/5$.
Apparently, for small \stepheights, the position of the maximum of the \PDF\
and the width of the central core are only weakly affected by the bandpass.

The dotted line in figure~\ref{fig:pdfFidBroadband} shows the limit to which
all curves converge for very large \stepheights.
For such large \stepheights, that is for~$k\param \gg 1$, the
sine and cosine functions in equation~(\ref{eq:intensitySH})
containing~$k\param$ vary rapidly with~$k$. The terms
containing~$k\param$ can therefore be replaced by their average
values which are~$1/2$ for the two quadratic terms
and~$(1-\cos(ka\diffAngle))/2$ for the mixed term. The dotted curve is
then given by
\begin{equation}
  \p_{ k\param \gg 1}(\diffAngle) =
        \frac{1}{(\pi ka)^{2}} \; \frac{1}{k_{2} - k_{1}} \;
            \int_{k_{1}}^{k_{2}} {\rm d}k \, \frac{1}{\diffAngle^{2}} \,
              [1 - \cos(\centerSegment k\diffAngle) \cos(k\diffAngle a)]
  \quad,
  \label{eq:intIntensityLargeDelta}
\end{equation}
where~$k_{1}$ and~$k_{2}$ are the integration limits.

 \begin{figure}[h]
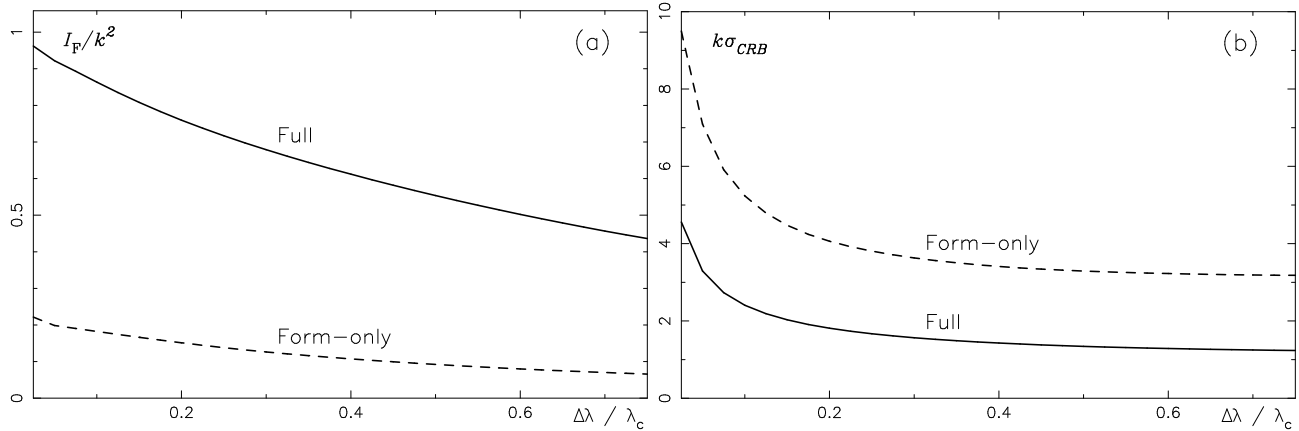

  \centerline{\hbox{
    \psfig{figure=efiBandpass.ps,width=85mm}
    \psfig{figure=errorBandpass.ps,width=85mm}}}
    \caption{\label{fig:efiErrorBandpass} {\small
          {\bf Shack-Hartmann sensor:}
          (a) Normalised Fisher information $I_{{\rm F}}/k^{2}$ 
          and (b) normalised \Cramer\ bound $\sigma_{{\rm CRB}}$
          as functions of the ratio of the width $\Delta \lambda$
          of the bandpass to the central wavelength $\lambda_{{\rm c}}$.
          {\it Solid}: full information,
          {\it dashed}: form-only information.}}
  \end{figure}
Figure~\ref{fig:pdfFidBroadband}b shows the
curves for the \FIDs\ for the same bandpasses as in figure~\ref{fig:pdfFidBroadband}a,
but for $\param=0$. In the ideal case the peaks in the \FID\ are,
according to equation~(\ref{eq:fisherInformationGeneral}), due to the
zeroes or small values of the \PDF\ at those locations. The peaks are now reduced
because the zeroes and the small values of the \PDFs\ are washed out
by the averaging process.
The \FIDs\ around the second maxima in the curves for $\param=0$ disappear
nearly completely for larger bandpasses,
and the \FID\ is therefore concentrated closer to the origin.

Figure~\ref{fig:efiErrorBandpass}a shows that, due to the weak dependences
of the form of the core and the position of the maximum on the bandpass,
the \EFI\ per photon as a function of the bandpass
decreases only weakly with the bandpass.
This is true for the full information (solid line) as well as
for the form-only information (dashed line). See section~\ref{sec:idealSH}
for the definition of form.
The decrease of the \EFI\ per photon with an increase of the bandpass is slower
than the proportional increase with the number of photons. Therefore the
\Cramer\ bound $\sigma_{{\rm CRB}}$ of the error in the estimate of the \stepheight\
decreases continuously with an increase of the bandpass.
This can also be seen in figure~\ref{fig:efiErrorBandpass}b which
shows $\sigma_{{\rm CRB}}$ in units of $\lambda/(2\pi)$
for one photon per nanometer of the bandpass.
 \begin{figure}[h]
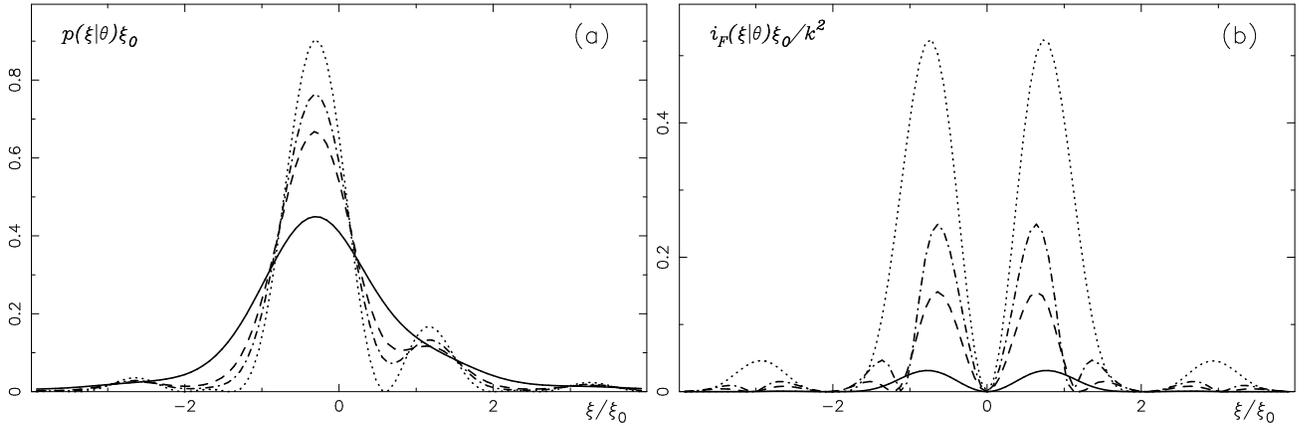

  \centerline{\hbox{
    \psfig{figure=pdfAtmosphere.ps,width=85mm}
    \psfig{figure=fidAtmosphere.ps,width=85mm}}}
    \caption{\label{fig:pdfFidAtmosphere} {\small
         {\bf Shack-Hartmann sensor:}
         (a) Likelihood functions for a \stepheight\ $\param = 0.1 \lambda$
             under different seeing conditions,
         (b) corresponding normalised \FIDs\ for $\param = 0$.
         Seeing conditions:
         {\it Dotted}: no atmosphere,
         {\it Dashed-dotted}: $r_{0}/2a = 2.25\,$,
         {\it Dashed}: $r_{0}/2a = 1.5\,$,
         {\it Solid}: $r_{0}/2a = 0.75\,$.}}
  \end{figure}
\subsection{Effects of atmospheric turbulence}
\label{sec:atmosphere}
Turbulence and temperature inhomogeneities in the atmosphere lead
to a non-planarity of the incoming wavefront.
These disturbances reduce the \EFI, and
consequently reduce the precision with which the \stepheight\
can be estimated.

The effects of the atmosphere can be taken into account by
writing the complex amplitude~$\complAmpl(x)$ in equation~(\ref{eq:pdfSHGeneral})
as a product
of the complex amplitude $\complAmpl_{{\rm s}}(x)$ related to the \boxheight\ and
a stochastic function~$U_{{\rm atm}}(x)$ describing the disturbance introduced by the
atmosphere. The likelihood function in equation~(\ref{eq:I1}) can then be written as
\begin{equation}
  \p(\diffAngle\cond\param)
  = \frac{k}4\pi a{} \; \int_{-a}^{+a}
      \, {\rm d}\varPupil' \int_{-a-\varPupil}^{+a-\varPupil} {\rm d}\varPupil \,
      \complAmpl_{{\rm s}}^{\ast}(\varPupil') \complAmpl_{{\rm s}}(\varPupil' + \varPupil) \;
           <U^{\ast}_{{\rm atm}}(\varPupil'+\varPupil)U_{{\rm atm}}(\varPupil')>
             \; e^{ik\diffAngle \varPupil}
  \quad.
     \label{eq:I1Atm}
\end{equation}
The expression in the brackets is the atmospheric structure function.
For Kolmogorov turbulence and neglecting the effects of the inner and outer
scales it is given by
\begin{equation}
  <U_{{\rm atm}}(\varPupil'+\varPupil)U_{{\rm atm}}(\varPupil')> =
           \exp \left[-3.44\left(\frac{\varPupil}{r_{0}}\right)^{5/3} \right]
  \quad,
     \label{eq:structureFunction}
\end{equation}
where $r_{0}$ is the atmospheric coherence length. For a given wavelength
it can be shown that the likelihood function depends on $\diffAngle$ and
$\xio$ only through the ratio $\diffAngle/\xio$ and on $r_{0}$ and $a$
only through the ratio $r_{0}/a$.

Figure~\ref{fig:pdfFidAtmosphere}a shows for a
\stepheight\ of~$\param = 0.1\lambda$ the influence of the
atmosphere on the \PDF\ for four different seeing conditions,
expressed as ratios $r_{0}/(2a)$. As in the case of the
integration over a finite bandpass the location of the maximum
does not depend on the seeing conditions.
However, the second maxima are washed out and the
width of the central core increases strongly with deterioration of the seeing.

The \FIDs\ for $\param=0$ and four different seeing conditions are shown 
in figure~\ref{fig:pdfFidAtmosphere}b. As in the case of larger bandpasses and
for similar reasons as given in the corresponding section~\ref{sec:broadband},
under the influence of the atmosphere the \FID\ is more concentrated near the origin
than the \FID\ under ideal conditions.

 \begin{figure}[h]
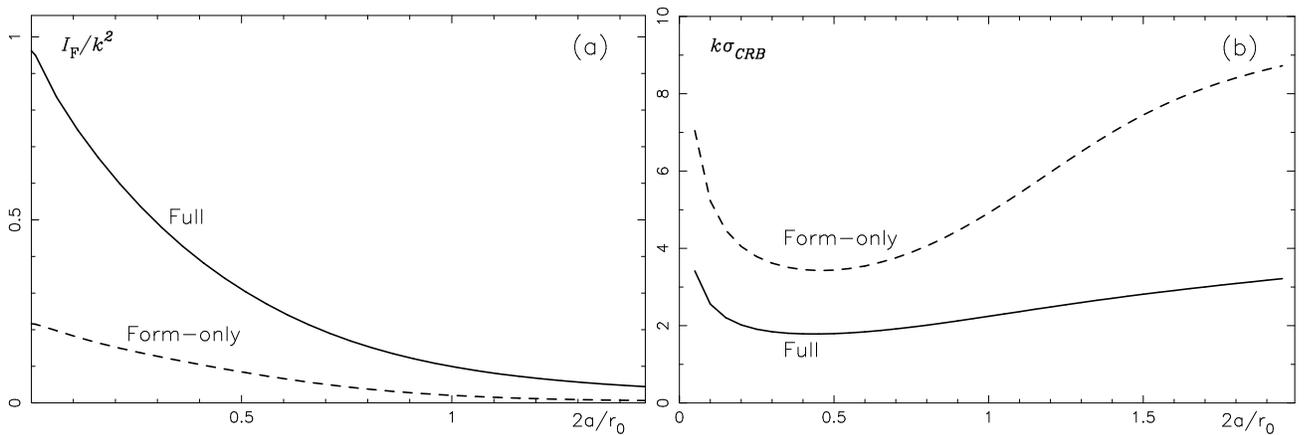

  \centerline{\hbox{
    \psfig{figure=efiAtmosphere.ps,width=85mm}
    \psfig{figure=errorAtmosphere.ps,width=85mm}}}
    \caption{\label{fig:efiErrorAtmosphere} {\small
          {\bf Shack-Hartmann sensor:}
          (a) Normalised Fisher information $I_{{\rm F}}/k^{2}$ 
          and (b) normalised \Cramer\ bound $\sigma_{{\rm CRB}}$
          as functions of the ratio of the aperture width $2a$
          to the atmospheric coherence length $r_{0}$.}}
  \end{figure}
For a design of the wavefront sensor one would usually assume a given seeing, that is
a given value for $r_{0}$, and then define the width of the lens.
Therefore, figure~\ref{fig:efiErrorAtmosphere}a shows
the normalised \EFI\ 
as a function of the ratio $2a/r_{0}$,
the full information as the solid line and the form-only information
as the dashed line.
The strong dependence of the width of the core on the seeing, mentioned above,
also explains the strong decrease of the \EFI\ with an
increase of the ratio $2a/r_{0}$.
If the diameter of the aperture is equal to $r_{0}$ the \EFI\ is already
reduced by a factor of approximately 10 compared to the \EFI\
under ideal conditions.

On the one hand the increase of the width of the lens increases the
Fisher information due to the larger number of photons. On the other
hand, the Fisher information per photon decreases rapidly with the size of the lens.
The combined effect is given by the product of
the Fisher information per photon times the width of the lens.
The corresponding curves for the \Cramer\ bounds in units of $\lambda/(2\pi)$
are plotted in figure~\ref{fig:efiErrorAtmosphere}b for
a flux density of one photon per millimeter of the subaperture.
Contrary to the situation for broadband measurements, the \Cramer\ bound
has a minimum, which is attained for a ratio $2a/r_{0} \approx 0.4$.
 \begin{figure}[h]
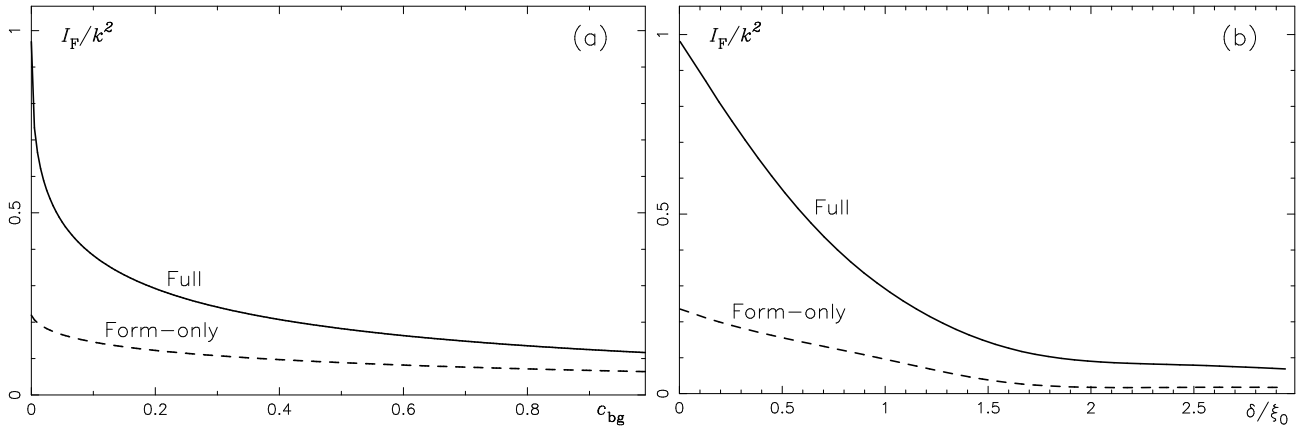

  \centerline{\hbox{
   \psfig{figure=efiBackground.ps,width=85mm}
   \psfig{figure=efiPixelWidth.ps,width=85mm}}}
   \caption{\label{fig:backgroundSampling}
     {\small
     {\bf Shack-Hartmann sensor:}
     (a) Normalised \EFI\ as a function of the readout flux in units of the
     flux from the sky source.
     (b) Normalised \EFI\ as a function of the pixel half-width $\delta$.}}
 \end{figure}
\subsubsection{Effects of detector readout noise}
The effect of the readout noise of the detector can be modelled by adding a constant
background~$c_{{\rm bg}}$ to the \PDF\ in equation~(\ref{eq:intensitySH}):
\begin{equation}
  p(\diffAngle\cond\param) = \frac{c_{{\rm bg}}}{\xio} + p_{0}(\diffAngle\cond\param)
                    \quad.
    \label{eq:pdfSHReadout}
\end{equation}
Here $p_{0}(\diffAngle\cond\param)$ is the \PDF\ for one photon without background noise.
The value of the constant $c_{{\rm bg}}$ is the ratio of the constant background
to the peak in the ideal likelihood function for $\param=0$.

Figure~\ref{fig:backgroundSampling}a shows
for a narrowband measurement the full \EFI\ (solid lines) and the
form-only \EFI\ (dashed lines) as functions of $c_{{\rm bg}}$.
The full \EFI\ declines initially much faster than the form-only \EFI.
The reason for this is that the zeroes in the \PDF, which are, according
to equation~(\ref{eq:fisherInformationGeneral}), responsible for the strong peaks
in the full \FID, disappear because of the constant background.
Since, for large backgrounds, the ratio of the full to the form-only \EFI\
only equals approximately two, the full \EFI\ is also more strongly affected than the
form-only \EFI\ for large backgrounds. Regarding figure~\ref{fig:backgroundSampling}b,
see next section.

 \begin{figure}[h]
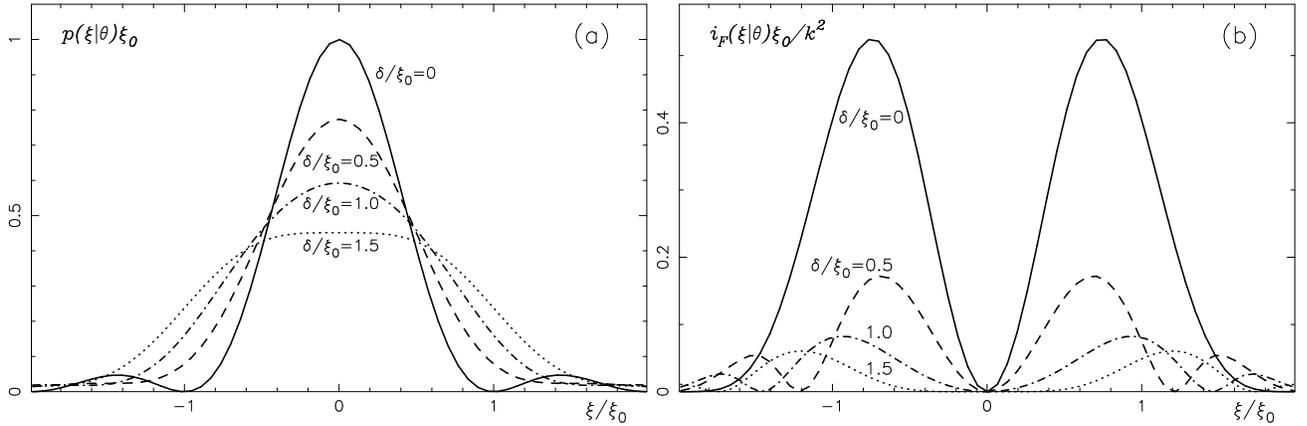

  \centerline{\hbox{
   \psfig{figure=pdfPixelWidth.ps,width=85mm}
   \psfig{figure=fidPixelWidth.ps,width=85mm}}}
   \caption{\label{fig:samplingPdfFid}
     {\small
     {\bf Shack-Hartmann sensor:}
     (a) Likelihood functions for pixel half-width of $\delta\rightarrow 0$ (solid),
      $\delta=0.5\xio$ (dashed), $\delta=\xio$ (dashed-dotted), and
      $\delta=1.5\xio$ (dotted).
     (b) Corresponding \FIDs.}}
 \end{figure}
\subsubsection{Effects of the detector pixelization}
The effect of finite pixel sizes $\delta$ can be modelled by
a product of a comb function, describing the positions of the centres
of the pixels, and a continuous \PDF\ which is obtained by averaging,
for each $\xi$, the initial \PDF\ over the
interval $[\xi-\delta/2,\xi+\delta/2]$.
Figure~\ref{fig:samplingPdfFid}a shows
the \PDF s for $\delta\rightarrow 0$ (solid), $\delta=0.5\xio$ (dashed),
$\delta=\xio$ (dashed-dotted), and $\delta=1.5\xio$ (dotted).
The corresponding \FIDs\ are shown in
figure~\ref{fig:samplingPdfFid}b for the full information.
The strong decline of the \FIDs\ with the pixel size is also apparent from
figure~\ref{fig:backgroundSampling}b, which shows that the decrease is
linear for small pixel widths.
The effect of the finite pixels size is similar for the full \EFI\
and the form-only \EFI.

\section{Conclusions}
This paper has shown that the quantum statistical information
relating to a \boxheight\ depends quadratically
on the ratio of the segment width to the width of the aperture. The maximum,
given by the square of the wavenumber per photon passing through the aperture, is attained
for the case where the segment covers half of the aperture.
In this special case the limiting precision for the measurement of the
\boxheight\ error is equal $\lambda/(2\pi)$, where $\lambda$
is the wavelength of the light. 
Under ideal conditions, that is with narrowband measurements
and negligible atmospheric or detector noise, both the Shack-Hartmann
and the Foucault sensor can, for small ratios of the segment size to the diameter
of the full aperture, potentially extract approximately
half of the ultimate quantum statistical information.
For the special case that the aperture is only filled by two segments with the phase step
at the centre of the aperture, the Shack-Hartmann device even reaches the ultimate precision,
and the precision obtained by the Foucault sensor is smaller by a factor of $\sqrt{2}$.
The other two sensors, Mach-Zehnder and curvature, require in general
approximately ten to twenty times the number of photons in order
to reach the same precision as the first
two sensors. In the case of the Shack-Hartmann sensor, broadband measurements always
increase the precision, whereas the optimum width for the subaperture corresponding to
one lenslet in the presence of atmospheric disturbances is approximately equal to half of
the atmospheric coherence length.
Also detector readout noise and the effects of a finite sampling
strongly reduce the precision. The effects of these disturbances on the other sensors may
be different from the effects on the Shack-Hartmann sensor
and remain to be investigated.

{\large {\bf Acknowledgements}}\\
The authors would like to thank R.N. Wilson, R. Mazzoleni, and N. Yaitskova for useful suggestions
on the manuscript and a critical reading of the final version.

\renewcommand{\theequation}{A-\arabic{equation}}
\setcounter{equation}{0}  
\section*{Appendix A : Expressions for the \PDF\ and \FID\ for the four sensors}
In this appendix the expressions for the \PDFs\ and \FIDs\ for the four sensors discussed
in the main text are derived.
\subsection*{A1: Shack-Hartmann sensor}
For monochromatic light with wavelength~$\lambda$ the Fraunhofer diffraction pattern
in the focal plane of an infinitely long cylindrical
lens covering a one-dimensional subaperture of width
$2a$ on the mirror can be calculated with a Fourier transformation.
The diffraction pattern may be viewed as a \PDF~$p(\diffAngle\cond\param)$ for
the detection of a photon at a coordinate $\xi$ in the focal plane.
$\xi$ is a unitless position variable, related to a true position
variable $\varDet$ in the focal or detector plane by $\xi = \varDet/f$, where $f$ is the
focal length of the lens.
If $\varPupil$ and~$\varPupil'$ are coordinates perpendicular to the axis
of the cylindrical lens,
and~$\complAmpl$ the complex field in the plane of the cylindrical lens,
the \PDF\ is given by
\begin{eqnarray}
  p(\diffAngle\cond\param)
  & = & \frac{k}{2\pi} \,
       <\int_{-a}^{+a} {\rm d}\varPupil
          \complAmpl^{\ast}(\varPupil'') e^{-ik\diffAngle \varPupil''} \;
         \int_{-a}^{+a} {\rm d}\varPupil' \complAmpl(\varPupil') e^{ik\diffAngle \varPupil'}>\\
  & = & \frac{k}{2\pi} \,
           \int_{-a}^{+a} {\rm d}\varPupil' \int_{-a}^{+a} {\rm d}\varPupil \;
                  <\complAmpl^{\ast}(\varPupil'')\complAmpl(\varPupil')>
                    e^{ik\diffAngle(\varPupil'-\varPupil'')}
  \quad.
    \label{eq:pdfSHGeneral}
\end{eqnarray}
The bracket denotes a statistical average over a possibly stochastic ensemble of
complex amplitudes $\complAmpl$. In this paper such a stochastic ensemble is used for the
treatment of atmospheric effects in section~\ref{sec:atmosphere}.

After the substitution~$\varPupil = \varPupil' - \varPupil''$ one gets
\begin{equation}
  \p(\diffAngle\cond\param)
  = \frac{k}{2\pi} \,
        \int_{-a}^{+a} \, {\rm d}\varPupil' \int_{-a-\varPupil}^{+a-\varPupil}\,
                 {\rm d}\varPupil
           <\complAmpl^{\ast}(\varPupil')\complAmpl(\varPupil' + \varPupil)>\;
                     e^{ik\diffAngle \varPupil}
     \quad.
     \label{eq:I1}
\end{equation}
Next, the non-stochastic wave function
$\complAmpl_{{\rm s}}$ related to the \boxheight\ as defined in equation~(\ref{eq:defPsiS}),
is introduced for $\complAmpl$ in equation~(\ref{eq:I1}). With $\xio=\lambda/(2a)$
as the location of the first zero in the diffraction pattern for $\param=0$ one gets
\begin{eqnarray}
  \p(\diffAngle\cond\param)
  & = & \frac{1}{\xio}\, \frac{1}{(\pi\diffAngle/\xio)^{2}}\;
    \left[ \sin^{2}(\pi\diffAngle/\xio) +
        2\sin^{2}(\pi \frac{\halfSegWidth}{a}\diffAngle/\xio)
              (1-\cos(2\pi\param/\lambda)) \right. \nonumber \\
    & & \left.
       + 2\sin(\pi\diffAngle/\xio)\sin(\pi \frac{\halfSegWidth}{a}\diffAngle/\xio)
       \left(\cos(\pi\frac{\centerSegment}{a}\diffAngle/\xio - 2\pi\param/\lambda)
                    - \cos(\frac{\centerSegment}{a}\diffAngle/\xio)\right)
         \right]
    \quad.
    \label{eq:intensitySH}
\end{eqnarray}
The \EFI\ can easily be computed by introducing the amplitude $q(\diffAngle\cond\param)$,
calculated from equation~(\ref{eq:intensitySH})
following equation~(\ref{eq:defProbAmplitude}),
into equation~(\ref{eq:EFI}). The lengthy expression for arbitrary values of $\centerSegment$,
$\halfSegWidth$, and $\param$ is not quoted here. For the special case of
$\param \rightarrow 0$ one obtains the following much shorter expression:
\begin{equation}
  \varFID(\diffAngle\cond\param \rightarrow 0) =
      \frac{4}{\pi^{2}}\frac{k^{2}}{\xio}\frac{1}{(\diffAngle/\xio)^{2}}\,
      \sin^{2}(\pi\frac{\halfSegWidth}{a}\diffAngle/\xio)
      \sin^{2}(\pi\frac{\centerSegment}{a}\diffAngle/\xio)
    \quad.
  \label{eq:fidSHSingle}
\end{equation}
\subsection*{A2: Mach-Zehnder sensor}
The amplitude $\complAmpl_{1}$ in the first, unmodified  beam in figure~\ref{fig:principleMZ}
is given by equation~({\ref{eq:defPsiS}).
After the light in the second beam passed through the phase plate and
the spatial filter, the amplitude $\complAmpl_{2}$ is given by the convolution
\begin{equation}
  \complAmpl_{2}(x) = \frac{1}{\sqrt{\lambda}}\int_{-\infty}^{\infty} {\rm d}x' \,
                  \complAmpl_{1}(x')T(x-x')
       \quad ,
             \label{eq:convolution}
\end{equation}
where $T(x)$ is the inverse Fourier transform of the transmission function
(\ref{eq:shapePinhole}) of the pinhole.
Since the detectors are in image planes of the pupil, the same coordinates
can be used as for the pupil, that is $\varDet=\varPupil$. 
The \PDFs\ on the two detectors are then
\begin{equation}
  p_{1,2}(x\cond\param) = \frac{1}{2}\, \{
     |\complAmpl_{1}(x\cond\param)|^{2} + |\complAmpl_{2}(x\cond\param)|^{2}
      \pm 2\, {\rm Re}[\complAmpl_{1}^{\ast}(x\cond\param)
           \complAmpl_{2}(x\cond\param)e^{i\phi}] \} \quad,
    \label{eq:pdfMZ}
\end{equation}
where the plus sign refers to $p_{1}$ and the minus sign to $p_{2}$.
With the special choice of the phase shift of $\phi=\pi/2$ and the
abbreviations
\begin{eqnarray}
   b & = & k\zeta\\
   \Psi_{{\rm L}} & = & \Phi(-b(a+x),b(\centerSegment-\halfSegWidth-x))\\   
   \Psi_{{\rm M}} & = & \Phi(b(\centerSegment-\halfSegWidth-x),
                             b(\centerSegment+\halfSegWidth-x))\\
   \Psi_{{\rm H}} & = & \Phi(b(\centerSegment+\halfSegWidth-x),b(a-x))\\
   \Phi(u,v) & = & \left\{ \begin{array}{ll}
          (2/\pi)\, \int_{u}^{v} \, {\rm d}z \,
             \sin z \, / \, z  & {\rm sharp} \; {\rm  edge} \; {\rm pinhole}\\
          (2/\sqrt{\pi}) \, \int_{u}^{v} \, {\rm d}z \, \exp(-z^{2}) 
                        & {\rm Gaussian} \; {\rm pinhole}
        \end{array} \right\}
      \quad ,
\end{eqnarray}
one obtains for the two \PDFs\
\begin{eqnarray}
  p_{1,2}(x\cond\param) & = & \frac{1}{8a} \, \{
     W(x,a,-a) + \frac{1}{4} \left[(\Psi_{{\rm L}}+\Psi_{{\rm H}})^{2}
                            + \Psi_{{\rm M}}^{2} 
              + 2\cos(k\param)(\Psi_{{\rm L}}+\Psi_{{\rm H}})\Psi_{{\rm M}}
                                \right] \nonumber\\
      & & \pm \, \sin(k\param) \; [
                    W(x,\centerSegment-\halfSegWidth,\centerSegment+\halfSegWidth)
                       (\Psi_{{\rm L}}+\Psi_{{\rm H}}) \nonumber \\
      & & \hspace*{17mm}
             - (W(x,-a,\centerSegment-\halfSegWidth) + W(x,\centerSegment+\halfSegWidth,a))
                           \Psi_{{\rm M}}
                ] \}
    \label{eq:pdfMZ1}
\end{eqnarray}
One of the assumptions for the derivation of equation~(\ref{eq:classicalFisherInfo})
was that the integrated intensity is independent of $\param$. In the case of the
Mach-Zehnder interferometer this is strictly only true for $\param \rightarrow 0$.
For both types of pinholes
the integrated intensity is a symmetric function of $\param$ with, according to numerical
calculations, a quasi parabolic behaviour around $\param=0$. Around the maximum at
$\param=0$ the derivative of the integrated intensity with respect to $\param$ is therefore
zero.

The \EFI\ is identical in the two beams behind the second beamsplitter.
With the definition
\begin{equation}
   \Psi^{2} = \Psi^{2}_{{\rm L}} + \Psi^{2}_{{\rm M}} + \Psi^{2}_{{\rm H}}
      \quad , 
      \label{eq:defPsiSquareMZ}
\end{equation}
the total \EFI\ as a sum of the \EFI\ from both beams behind the second
beamsplitter is, in the limit $\param \rightarrow 0$, given by
\begin{equation}
I_{{\rm F,MZ}} = \frac{k^{2}}{4a} \, {\big [}
            \int_{-a}^{+a} \, {\rm d}x \,
            \Psi^{2}_{{\rm M}} / \left(1 + \Psi^{2}/4 \right)
           + \int_{\centerSegment-\halfSegWidth}^{\centerSegment+\halfSegWidth} \, {\rm d}x \,
            \left((\Psi^{2}_{{\rm L}} + \Psi^{2}_{{\rm H}})^{2}
                      - \Psi^{2}_{{\rm M}} \right) / \left(1 + \Psi^{2}/4 \right)
               {\big ]} \quad .
    \label{eq:efiMZ}
\end{equation}
\subsection*{A3: Foucault sensor}
The Fourier transform $T(x)$ of a Heaviside step function representing the effect of
the knife edge in the focal plane is given by
\begin{equation}
  T(x) = \frac{\sqrt{\lambda}}{2\pi}\, \left(\pm\frac{1}{ix} + \pi \delta(x) \right)
         \quad ,
             \label{eq:FTStepFunction}
\end{equation}
where the signs $+$ and $-$ refer to the two channels with the transmission equal to zero
for $\xi<0$ and $\xi>0$, respectively. Introducing equation~(\ref{eq:FTStepFunction})
into equation~(\ref{eq:convolution}) gives with the definitions
\begin{eqnarray}
  Y_{1} & = & \ln\frac{(\centerSegment + \halfSegWidth - x)^{2}}
                          {(\centerSegment - \halfSegWidth - x)^{2}} \\
  Y_{2} & = & \ln\frac{(a-x)^{2}}{(a+x)^{2}}
\end{eqnarray}
the expression for the likelihood function on the two detectors
\begin{eqnarray}
  |\complAmpl_{2,{\rm FC}}(x\cond\param)|^{2} & = &
      \frac{1}{8a\pi^{2}} \; \{\frac{1}{4}Y_{1}^{2} + \pi^{2}W(x,-a,a)
         + \frac{1}{2}(1-\cos(k\param)) \, Y_{2}(Y_{2} - Y_{1}) \nonumber\\
       & &  \hspace{10mm} \pm \sin(k\param) \pi (W(x,-a,a)Y_{2} -
         W(x,\centerSegment-\halfSegWidth,\centerSegment+\halfSegWidth)Y_{1}) \}
         \quad.
             \label{eq:psi2FC}
\end{eqnarray}
The inversion of the sign in front of the sine-term can be understood
from the fact that the intensity distribution must be symmetric with respect to
a change of the transmission from $\xi<0$ to $\xi>0$ and a simultaneous
change of the sign of the \stepheight~$\param$.
A similar expression for the centred single-border configuration has been derived
in~\cite{Pinna2004}.

For one channel alone the requirement for the integrated intensity to be independent
of $\param$ is not fulfilled, since the centre of gravity of the diffraction pattern
in the focal plane depends linearly on $\param$ around $\param=0$, as shown by
equation~(\ref{eq:shiftCOGSmall}). However, the requirement is obviously fulfilled for the
sum of the integrated intensities in the two channels. Therefore, the formalism has
to use the sum of the \FIDs\ in the two channels, which are, however,
in the limit $\param \rightarrow 0$, independent of the channel.
For $\param \rightarrow 0$ the sum of the \EFI\ in the two channels is given by
\begin{eqnarray}
  I_{{\rm F,FC}} = \frac{k^{2}}{4a} \int_{-a}^{+a}
     \left(Y_{2} -
       W(x,\centerSegment-\halfSegWidth,\centerSegment-\halfSegWidth)Y_{1} \right)^{2}
        {\big /} \left(\pi^{2} + Y_{1}^{2}/4 \right)
\end{eqnarray}
\subsection*{A4: Curvature sensor}
The one-dimensional equivalent of the two-dimensional expression for the complex
amplitude in the plane of the detector given in \cite{Roddier1987}
\cite{Gonzales2001} is
\begin{equation}
  \complAmpl_{{\rm CV}}(x \cond \param) =
        \frac{1}{\sqrt{\lambda (f-l)}} e^{ikf}e^{i\pi x^{2}/(\lambda f)}
     \, \int_{-a}^{+a} {\rm d}\varPupil \, \complAmpl(\varPupil \cond \param) \,
       e^{i\frac{\pi}{2} (\varPupil - fx/l)^{2} / \rho^{2}}
     \quad ,
          \label{eq:psiCv1Dim}
\end{equation}
where $U$ is the complex amplitude in the pupil and
\begin{equation}
   \rho = \sqrt{\frac{\lambda f (f-l)}{2l}}.
\end{equation}
With the definition
\begin{equation}
  \eta = (\varPupil - \frac{f}{l}x) / \rho
\end{equation}
equation~(\ref{eq:psiCv1Dim}) can be written as
\begin{equation}
  \complAmpl_{{\rm CV}}(x\cond \param) = \sqrt{\frac{i}{2}}\, \sqrt{\frac{f}{l}} \, e^{ikf}\,
               e^{i\pi x^{2}/(\lambda f)}
     \, \int_{(-a-fx/l)/\rho}^{(+a-fx/l)/\rho} {\rm d}\eta \; U_{{\rm s}}(\eta \cond \param) \;
       e^{i\pi \eta^{2}/2}  \quad .
          \label{eq:psiCv}
\end{equation}
The expression for $\complAmpl_{{\rm CV}}$ has been normalised such that
\begin{equation}
  \int_{\infty}^{+\infty} {\rm d}x \, |\complAmpl_{{\rm CV}}(x\cond \param)|^{2} = 1
     \label{eq:normalisationPsiCV}
\end{equation}
Introducing $U_{{\rm s}}$ from equation~(\ref{eq:defPsiS}) for $\complAmpl$ 
in equation~(\ref{eq:psiCv}) one obtains with the abbreviations
\begin{equation}
   \begin{array}{llllll}
      C_{1} & = & {\cal C}((-a - fx/l)/\rho) &
      S_{1} & = & {\cal S}((-a - fx/l)/\rho \\
      C_{2} & = & {\cal C}((\centerSegment - \halfSegWidth - fx/l)/\rho) &
      S_{2} & = & {\cal S}((\centerSegment - \halfSegWidth - fx/l)/\rho \\
      C_{3} & = & {\cal C}((\centerSegment + \halfSegWidth - fx/l)/\rho) &
      S_{3} & = & {\cal S}((\centerSegment + \halfSegWidth - fx/l)/\rho \\
      C_{4} & = & {\cal C}((a - fx/l)/\rho) &
      S_{4} & = & {\cal S}((a - fx/l)/\rho \\
      C_{ij} & = & C_{j} - C_{i} &
      S_{ij} & = & S_{i} - S_{j}\\
   \end{array}
    \quad,
\end{equation}
where ${\cal C}$ and ${\cal S}$ denote the Fresnel integrals,
the following expression for the likelihood function:
\begin{eqnarray}
  |\complAmpl_{{\rm CV}}(x\cond \param)|^{2} & = & \frac{f}{4al} \; \huge{(}
      C_{21}^{2} + S_{21}^{2} + C_{32}^{2} + S_{32}^{2} + C_{43}^{2} + S_{43}^{2}
    \nonumber\\
    && \hspace{6mm} + 2\; \cos(k\param) \; [C_{21}C_{32} + S_{21}S_{32} + C_{32}C_{43} + S_{32}S_{43}]
    \nonumber\\ 
    && \hspace{6mm} + 2\; \sin(k\param) \; [S_{12}C_{32} - C_{21}S_{32} + S_{32}C_{43} - C_{32}C_{43}]
    \nonumber\\
    && \hspace{6mm} + 2C_{21}C_{43} + 2S_{21}S_{43} {\huge )}  \quad .
      \label{eq:pdfCV}
\end{eqnarray}
The corresponding expression for the Fisher information density can be readily calculated from
the equations~(\ref{eq:defFisherDensity}), and (\ref{eq:pdfCV}) and is not presented here.


\end{document}